\documentclass[lettersize]{IEEEtran}

\usepackage{booktabs}
\usepackage{graphicx}
\usepackage{subfigure}
\usepackage[table]{xcolor}
\usepackage{amsmath}
\usepackage{hyperref}
\usepackage{cleveref}
\usepackage{balance}
\usepackage{enumitem}
\usepackage{multirow}
\usepackage{xspace}
\usepackage{makecell}
\usepackage{tikz}
\usetikzlibrary{shapes, arrows.meta, positioning, fit, backgrounds, calc, decorations.pathreplacing}
\usepackage{pifont}
\usepackage{standalone}
\usepackage{listings}
\usepackage{tcolorbox}
\usepackage{url}

\usepackage{caption}
\newenvironment{packeditemize}{
	\begin{list}{$\bullet$}{
			\setlength{\labelwidth}{4pt}
			\setlength{\itemsep}{0pt}
			\setlength{\leftmargin}{\labelwidth}
			\addtolength{\leftmargin}{\labelsep}
			\setlength{\parindent}{0pt}
			\setlength{\listparindent}{\parindent}
			\setlength{\parsep}{0pt}
			\setlength{\topsep}{1pt}}}{\end{list}}

\newcommand{\ie}{i.e.,\xspace}
\newcommand{\eg}{e.g.,\xspace}
\newcommand{\etal}{\textit{et al.}\xspace}
\newcommand{\para}[1]{\smallskip\noindent\textbf{#1.}}
\newcommand{\cmark}{\textcolor{green!80!black}{\ding{51}}}
\newcommand{\xmark}{\textcolor{red}{\ding{55}}}
\newcommand{\pmark}{\textcolor{blue!90}{\ding{109}}}
\newcommand{\Uzero}{\ensuremath{\mathbf{U_0}}\xspace}
\newcommand{\Uone}{\ensuremath{\mathbf{U_1}}\xspace}
\newcommand{\Utwo}{\ensuremath{\mathbf{U_2}}\xspace}

\newtcolorbox{takeaway}[1][]{%
  colback=gray!8, colframe=gray!50, fonttitle=\bfseries,
  title=Takeaway, #1, boxrule=0.5pt, arc=2pt}

\definecolor{bgpurple}{HTML}{E8E0F0}
\definecolor{bgblue}{HTML}{D8E4F0}
\definecolor{bggreen}{HTML}{E0F0E0}
\definecolor{bgred}{HTML}{F0E0E0}
\definecolor{bgyellow}{HTML}{F5F0D0}
\definecolor{nodecyan}{HTML}{66CCEE}
\definecolor{nodeorange}{HTML}{EE8866}
\definecolor{nodepink}{HTML}{EE6677}
\definecolor{nodegreen}{HTML}{228833}
\definecolor{nodeblue}{HTML}{4477AA}
\definecolor{nodegrey}{HTML}{BBBBBB}
\definecolor{arrowred}{HTML}{CC3333}
\definecolor{arrowblue}{HTML}{3355AA}
\definecolor{arroworange}{HTML}{DD7700}
\definecolor{textdark}{HTML}{333333}

\graphicspath{{../figures/}}

\begin{document}

\title{In the Margins: An Empirical Study of \\Ethereum Inscriptions}

\author{
\IEEEauthorblockN{Xihan Xiong$^1$, Minfeng Qi$^2$, Shiping Chen$^5$, Guangsheng Yu$^3$, Zhipeng Wang$^4$, Qin Wang$^{3,5}$} \\
\textit{$^1$Imperial College London} $|$  \textit{$^2$City University of Macau}  $|$ \textit{$^3$University of Technology Sydney}   \\  \textit{$^4$The University of Manchester} $|$ \textit{$^5$CSIRO Data61} 

}

\maketitle

\begin{abstract}

Ethereum Inscriptions (Ethscriptions) repurpose Ethereum calldata into a persistent inscription channel by embedding \texttt{data:}~URI payloads. 
These transactions typically target externally owned accounts, allowing the payload to bypass EVM execution while remaining permanently replicated across full nodes. 
Although calldata was originally designed for compact smart-contract parameters, this repurposing enables structured data embedding with long-term storage consequences.

We present the first large-scale empirical study of Ethscriptions, treating them as a distinct \emph{calldata-resident workload} rather than merely a subset of general calldata usage.
Our analysis focuses on the \textit{Ethscription} operational subset, which consists of payloads that decode to JSON and conform to a token-operation grammar (e.g., \texttt{p}, \texttt{op}, \texttt{tick}, \texttt{amt}).
From $6.27$ million Ethscription candidates (\Uone), we extract $4.75$ million Ethscription operations (\Utwo, $75.8\%$ of \Uone). 
This result shows that structured token-like activity dominates the ecosystem. 
Our measurements further reveal (i) a complete workload lifecycle compressed into nine months (bootstrap, expansion, saturation), (ii) proliferation of $30$+ competing protocols without convergence toward a dominant standard, (iii) a lifecycle funnel exhibiting $201\times$ deploy-to-mint amplification and a $57.6{:}1$ mint-to-transfer collapse indicative of speculative minting, (iv) extreme participation inequality (Gini~$0.86$), and (v) a measurable permanent data footprint imposed on the Ethereum network.

\end{abstract}

\begin{IEEEkeywords}
Blockchain, Ethereum, Inscriptions, Ethscriptions, Calldata, Empirical measurement, JSON token protocols
\end{IEEEkeywords}

\section{Introduction}
\label{sec:intro}

Ethereum calldata was designed as an input field for smart-contract execution. It carries function selectors and ABI-encoded parameters that the EVM reads during transaction processing~\cite{wood2014ethereum, buterin2014whitepaper}. Accordingly, the gas pricing model and execution semantics treat calldata as transient execution input intended to support computation rather than long-term data storage.  In practice, however, permissionless blockchain systems allow users to repurpose protocol mechanisms beyond their original design assumptions. 
Ethscriptions~\cite{lehman2023ethscriptions}, introduced in June 2023, follow this pattern by embedding \texttt{data:} URI payloads in calldata, thereby using an execution input field as an inscription channel. These transactions typically target externally-owned accounts (EOAs), so the EVM does not interpret the payload, while the data still becomes part of the permanent blockchain record. In effect, calldata is used as replicated storage, with its contents retained by all full nodes.

This practice leads to a class of transactions whose primary goal is durable data embedding rather than computation or value transfer. We call this a \emph{payload-resident workload}. The idea is conceptually similar to Bitcoin Ordinals~\cite{rodarmor2023ordinals} and BRC-20~\cite{domo2023brc20}, which attach data to witness fields, but Ethscriptions rely on Ethereum calldata and therefore have different economic and architectural implications. Calldata fees are set to reflect execution bandwidth rather than long-term archival cost, yet inscription payloads impose permanent storage burden shared by the network. This disconnect turns an individual transaction choice into a network-wide externality, and it raises systems questions about pricing, incentives for storage, and protocol-level resource allocation.

\begin{figure}[!t]
    \centering
    \includegraphics[width=\linewidth]{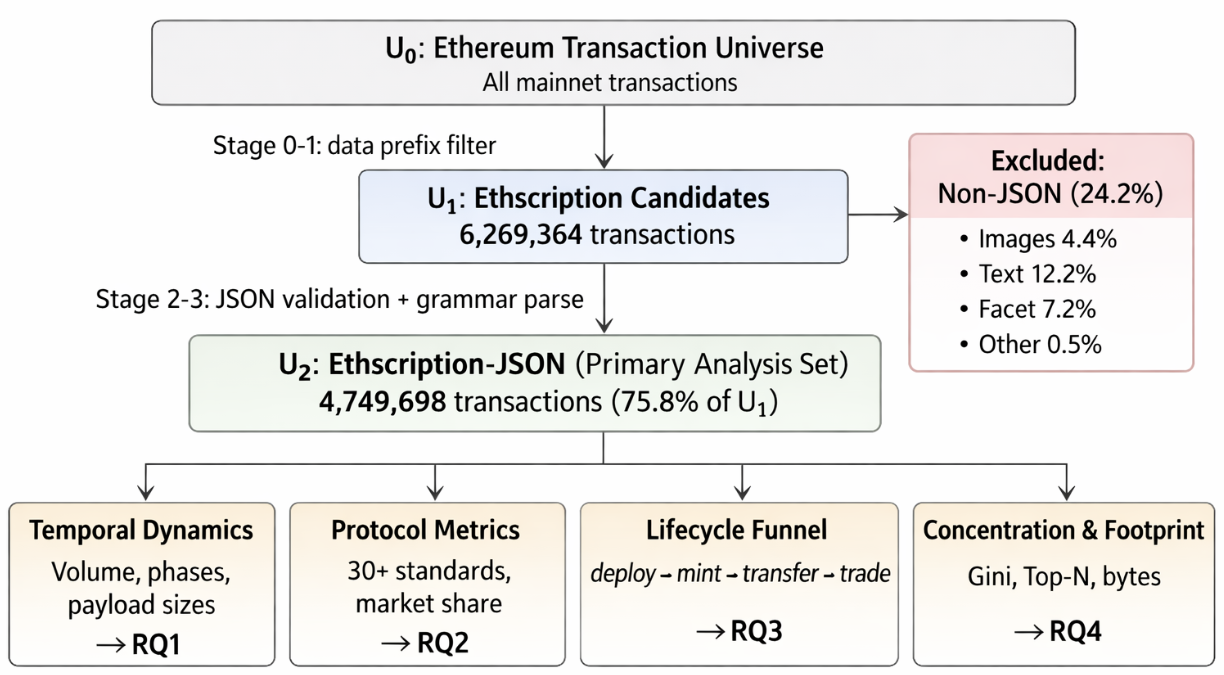}
    \caption{Processing pipeline and dataset hierarchy. Raw Ethereum transactions (\Uzero) are filtered by \texttt{data:}prefix to yield Ethscription candidates (\Uone=6.27M). MIME classification establishes composition context. JSON validation and grammar parsing extract the primary analysis set (\Utwo=4.75M, 75.8\% of \Uone). All core measurements are computed on \Utwo.}
    \label{fig:methodology}
\end{figure}

Despite rapid adoption, Ethscriptions are not well characterized from a measurement standpoint. Public discussions are often anecdotal or market-driven, and prior measurement work typically treats calldata as a single category, mixing rollup data availability, contract inputs, and arbitrary payload embedding~\cite{wood2014ethereum}. Without separating inscription workloads from other uses, it is difficult to analyze their dynamics, participation structure, or long-run infrastructure impact. We therefore study Ethscriptions as a distinct, measurable system phenomenon.

This paper is \textbf{not} a study of general Ethereum calldata usage. We exclude rollup data availability workloads, ABI-encoded contract inputs, and other calldata-heavy applications. Instead, we define a precise workload boundary using the \texttt{data:} URI prefix and analyze only Ethscription candidates. Within this boundary, Ethscriptions are operationally identified by structured JSON payloads that encode token-like actions interpreted by off-chain indexers. A small fraction of \texttt{data:}-prefixed transactions contains media or free-form content; we treat these as ecosystem context rather than part of the operational workload. Accordingly, our main analysis focuses on transactions whose payloads satisfy the Ethscription operation grammar, while non-operational payloads are reported only to describe overall composition. All statistics explicitly distinguish scope: \Uone denotes all \texttt{data:}-prefixed candidates, and \Utwo denotes the subset that conforms to the operational grammar used by indexers. This distinction reflects measurement scope rather than a different protocol category.

Beyond the workload itself, Ethscriptions highlight a broader architectural pattern. Consensus provides data availability and ordering, while semantic meaning is assigned by off-chain indexers that reconstruct state by parsing and interpreting payloads. The resulting semantic layer is software-defined: correctness depends on indexer behavior rather than protocol enforcement. This motivates a measurement-first study that begins with what is being submitted on-chain.

We pose five progressive research questions (RQs):

\begin{itemize}
\item\textbf{RQ0:}  What is an Ethscription at the transaction/payload level, and what is the Ethscription operation grammar?
\item \textbf{RQ1:}  How does the Ethscription workload evolve over time (volume, phases, payload sizes)?
\item \textbf{RQ2:}  What protocol identifiers and operation pipelines emerge from Ethscription payloads, and what are their lifecycle conversion properties?
\item \textbf{RQ3:}  Who generates the Ethscription workload, and what persistent data footprint does it impose on the network?
\item \textbf{RQ4:}  What design lessons follow for calldata pricing, data-channel separation, and indexer-defined semantics?
\end{itemize}

\para{Contributions}
This paper makes four contributions:

\begin{packeditemize}
\item[$\triangleright$] We construct a scope-precise Ethscription dataset with explicit boundaries, where \Uone contains $6.27$M \texttt{data:}-prefixed candidates and \Utwo contains $4.75$M JSON operations. We further provide a fully documented parsing pipeline that derives \Utwo from \Uone.

\item[$\triangleright$] We formalize the Ethscription operation grammar and introduce a normalized operation taxonomy covering more than 30 protocol identifiers and more than 20 operation types.

\item[$\triangleright$] We provide a comprehensive longitudinal measurement of the Ethscription workload, covering workload evolution, protocol growth, lifecycle conversion behavior, sender concentration, and persistent data footprint. 
All results are computed on \Utwo with explicit scope labeling.

\item[$\triangleright$] 
We analyze the system-level implications of Ethscriptions repurposing calldata into a permanent data channel. 
We focus on pricing incentives, data-channel separation, and the robustness limits of indexer-defined semantics.
\end{packeditemize}

The rest of the paper is organized as follows. \Cref{sec:background} reviews calldata economics and prior inscription designs across chains. \Cref{sec:technical} details the Ethscription transaction structure and JSON operation grammar. \Cref{sec:methodology} introduces the dataset hierarchy and measurement methodology. \Cref{sec:results} presents the main measurement results on \Utwo. \Cref{sec:discussion} discusses system-level implications, \Cref{sec:limitations} outlines limitations, and \Cref{sec:related} surveys related work.

\section{Technical Foundations}
\label{sec:background}

\subsection{Calldata Persistence and Cost Model}
\label{sec:calldata_cost}

Ethereum calldata~\cite{xiong2025toxic,xiong2025talking} is charged at $16$~gas per non-zero byte and $4$~gas per zero byte, in addition to the fixed base transaction cost of $21{,}000$~gas~\cite{wood2014ethereum}. This pricing scheme was designed for transmitting smart contract parameters, which are typically compact function calls consisting of a few hundred bytes. Unlike contract storage, which can be modified through \texttt{SSTORE} or partially reclaimed via mechanisms such as \texttt{SELFDESTRUCT}, calldata forms part of the immutable transaction payload. Once a transaction is confirmed, its calldata remains permanently stored across all full nodes~\cite{alzoubi2024blockchain}.

As an illustration, a typical $75$-byte JSON mint operation incurs roughly $75 \times 16 \approx 1{,}200$ gas for calldata (assuming all bytes are non-zero as an upper bound), in addition to the $21{,}000$~gas base transaction cost. At a gas price of $30$~Gwei, the total cost is approximately $22{,}200 \times 30\,\text{Gwei} \approx 6.66\times10^{-4}$~\textsc{ETH} (about $\$1.3$ assuming $\$2{,}000$/\textsc{ETH} ). Because fiat exchange rates fluctuate substantially over time, we primarily report costs in gas and byte-level units throughout the paper to maintain consistency across measurements.

EIP-4844~\cite{buterin2022eip4844} introduced ``blob'' transactions that provide temporary data availability with an approximately $18$-day retention window. This design targets workloads that rely on calldata permanence, particularly rollup data availability. Ethscriptions differ in that they use calldata for permanent data embedding associated with token-like operations rather than temporary execution-related data.

\subsection{Cross-chain Inscription Mechanisms}
\label{sec:cross_chain}


The inscription paradigm first emerged in the Bitcoin ecosystem through Ordinals~\cite{li2024bitcoin,rodarmor2023ordinals}, which associate serial numbers with individual satoshis and allow arbitrary data to be attached using Taproot witness fields~\cite{writingonwall2024,wang2023brc20} (\Cref{tab:comparison_brc20}). Because witness data is discounted relative to base transaction data, with fees reduced by approximately $75\%$, it became economically viable to embed application-level data directly on-chain. Building on this mechanism, BRC-20~\cite{domo2023brc20} introduced a JSON-based token protocol in which deploy, mint, and transfer operations are encoded as structured payloads stored in witness data. Token balances and state transitions are not enforced by the Bitcoin protocol itself but reconstructed by off-chain indexers that interpret these payloads.

Ethscriptions~\cite{lehman2023ethscriptions} adopt a similar design philosophy on Ethereum but rely on calldata rather than witness data as the embedding channel. Although both approaches use transaction data as a carrier and depend on off-chain interpretation, the underlying economic and architectural assumptions differ substantially. Bitcoin inscriptions exploit discounted witness storage originally introduced for scalability improvements, whereas Ethereum calldata pricing reflects execution bandwidth rather than archival cost. As a result, inscription workloads on Ethereum inherit a persistence model that was not explicitly designed for large-scale data embedding, leading to different cost incentives and infrastructure impacts (\Cref{tab:comparison_brc20}).

\begin{table}[t]
\centering
\caption{Ethscriptions (Ethereum) vs.\ BRC-20 (Bitcoin)}
\label{tab:comparison_brc20}
\small
\begin{tabular}{c|c|c}
\toprule
\multicolumn{1}{c|}{\textbf{Property}} & 
\multicolumn{1}{c}{\textbf{Ethscriptions}} & 
\multicolumn{1}{c}{\textbf{BRC-20~\cite{wang2023brc20}}} \\
\midrule
Data field & Calldata & Witness (Taproot) \\
\rowcolor{gray!10}
Fee model & 16 gas/byte & 1/4 weight discount \\
Smart contracts & Bypasses EVM & No smart contracts \\
\rowcolor{gray!10}
State tracking & Off-chain indexer & Off-chain indexer \\
Protocol variants & 30+ standards & BRC-20 dominant \\
\rowcolor{gray!10}
Payload format & \texttt{data:} URI + JSON & JSON in witness \\
Data permanence & Permanent & Permanent \\
\bottomrule
\end{tabular}
\end{table}

\section{Technical Dissection for Ethscriptions}
\label{sec:technical}

This section addresses \textbf{RQ0}: What is an Ethscription at the transaction/payload level, and what is the Ethscription operation grammar? We formalize the underlying structure and operational semantics that form the foundation of our analysis.

\subsection{What Is an Ethscription? (\Uone Definition)}
\label{sec:ethscription_def}
\noindent\textbf{Definition}. We define \Uone as the set of transactions whose calldata begins with the \texttt{data:} URI prefix defined in RFC~2397. In practice, most Ethscriptions follow a simple pattern: a zero-value transaction sent to an EOA, often the sender itself, with the payload embedded directly in the calldata field.


\begin{lstlisting}[
caption={Ethscription transaction format.},
label={lst:ethscription},
captionpos=t
]
Tx {
  to    = 0xRECIPIENT      // externally-owned account
  value = 0                // no ether transferred
  data  = 0x646174613a...  // hex("data:...")
}
\end{lstlisting}

\para{Execution semantics and gas cost}
Because these transactions target an EOA rather than a contract account, no bytecode is executed by the EVM. The gas cost therefore consists of the base transaction cost of $21{,}000$~gas plus the calldata charge ($16$~gas per non-zero byte and $4$~gas per zero byte)~\cite{wood2014ethereum}. There is no additional storage or contract execution overhead.

\para{Empirical patterns}
The \Uone dataset exhibits several consistent behavioral patterns that distinguish inscription-style activity from ordinary transaction flows. Approximately $70.7\%$ of transactions are self-sends (\texttt{from} = \texttt{to}), and $97.8\%$ carry zero \textsc{ETH} value, indicating that the primary purpose is payload publication rather than asset transfer. In addition, $14.4\%$ of transactions target the null address, suggesting that the recipient field is often incidental to the data-embedding intent. These transactions are typically minimal in value-transfer semantics but non-trivial in calldata size, reflecting their focus on byte-level data embedding under the prevailing gas model. Importantly, these properties are descriptive rather than definitional: they summarize common statistical characteristics of \Uone but are not required conditions for inclusion, which is determined solely by the presence of the \texttt{data:} URI prefix in calldata.

\subsection{From Ethscriptions to Ethscription (\Utwo Definition)}
\label{sec:u1_to_u2}

The \Uone set includes a wide range of payload types embedded through the \texttt{data:} URI mechanism. These payloads span images (PNG, SVG, JPEG, GIF, WebP), plain text, HTML fragments, Facet Protocol operations, and structured JSON token operations. \Cref{tab:composition} summarizes the detailed content-type distribution within \Uone.

\begin{table}[t]
\centering
\caption{Content-Type of \Uone (Ethscription Candidates)}
\label{tab:composition}
\small
\begin{tabular}{l|cc}
\toprule
\rowcolor{gray!10}
\multicolumn{1}{c}{\textbf{Content Type}} & \textbf{Count} & \textbf{Share (\%)} \\
\midrule
JSON (token protocols) $\rightarrow$ \Utwo & 4,749,698 & 75.8 \\
Short text (\texttt{data:,...}) & 683,115 & 10.9 \\
Facet Protocol & 449,909 & 7.2 \\
Images (PNG/SVG/JPEG/GIF/WebP) & 274,314 & 4.4 \\
Long text (\texttt{text/plain}) & 78,911 & 1.3 \\
HTML & 25,768 & 0.4 \\
Others & 7,649 & 0.1 \\
\midrule
\rowcolor{gray!10}
\multicolumn{1}{c|}{\textbf{Total \Uone}} & \textbf{6,269,364} & \textbf{100.0} \\
\bottomrule
\end{tabular}
\end{table}

Among these categories, JSON payloads form the dominant workload. The JSON subset (\Utwo) accounts for $75.8\%$ of all Ethscription transactions, indicating that most activity is associated with structured token operations rather than media-style inscriptions. For this reason, our measurement study focuses primarily on \Utwo. Non-JSON inscriptions (\Uone$\setminus$\Utwo) are excluded from the main analyses and are considered only when reporting overall ecosystem composition.

JSON payloads are particularly important because they encode structured operations such as \texttt{deploy}, \texttt{mint}, and \texttt{transfer}. These operations are interpreted by off-chain indexers, which reconstruct token balances and ownership state by parsing the payload content. In contrast to media inscriptions, whose meaning is primarily representational, JSON operations define executable protocol semantics within the indexing layer. Their compact size (with a median payload length of approximately 75 bytes) also makes individual operations inexpensive, which in turn supports high transaction volumes during minting phases.

\subsection{Ethscription Operation Model}
\label{sec:operation_model}

To support consistent measurement, we define an \emph{Ethscription operation model} that specifies how JSON payloads are identified, normalized, and interpreted as operational events. The goal of this model is to connect raw calldata payloads with the structured workload analyzed throughout the paper.





\subsubsection{Operation grammar}

An Ethscription operation is encoded as a JSON object that contains at least two fields. The field \texttt{p} specifies the protocol identifier (for example \texttt{erc-20}, \texttt{ierc-20}, or \texttt{rerc-20}), which we observe as lowercase alphanumeric strings that may include hyphens. The field \texttt{op} specifies the operation type, typically values such as \texttt{deploy}, \texttt{mint}, or \texttt{transfer}. 

Additional attributes may appear depending on the operation. These include \texttt{tick} for the token ticker, \texttt{amt} for the operation amount (encoded either as a string or an integer), \texttt{max} for the maximum supply specified during deployment, and \texttt{lim} for the per-mint limit. Some protocols introduce extra fields, such as \texttt{pow} for proof-of-work difficulty in \texttt{ierc-pow}, together with a corresponding \texttt{nonce}.





\subsubsection{Parsing and normalization}

Because Ethscription semantics are interpreted by off-chain indexers rather than enforced by the Ethereum protocol, normalization is required before measurement. Without normalization, minor syntactic differences across payloads could produce artificial protocol variants or distort operation counts. We therefore apply a deterministic normalization pipeline. Operation names are converted to lowercase and common aliases are merged (for example \texttt{Transfer} becomes \texttt{transfer}). Numeric attributes are parsed consistently regardless of encoding, so values expressed as strings (e.g., \texttt{"1000"}) and integers (\texttt{1000}) are treated identically. Leading and trailing whitespace is removed, URL-encoded payloads (such as \texttt{\%7B}) are decoded prior to JSON parsing, and duplicate keys follow a last-value-wins rule.

\subsubsection{Operational lifecycle model}




After normalization, operations in \Utwo\ can be organized into a four-stage lifecycle that describes how tokens are created, distributed, and exchanged. The first stage is \textit{deploy}, which defines a new token and specifies supply parameters such as \texttt{tick}, \texttt{max}, and \texttt{lim}. The second stage is \textit{mint}, where participants allocate or claim token units from an existing deployment using the fields \texttt{tick} and \texttt{amt}. The third stage is \textit{transfer}, which moves tokens between addresses and includes variants such as \texttt{transfer} and \texttt{proxy\_transfer}. The final stage is \textit{trade}, which captures marketplace interactions including operations such as \texttt{freeze\_sell}, \texttt{list}, \texttt{buy}, and \texttt{sell}. This lifecycle abstraction provides the analytical basis for the conversion and funnel measurements presented in Section~\ref{sec:methodology}.

\begin{lstlisting}[caption={Example of deploy and mint operations.},
label=lst:deploy_mint, captionpos=t]
// Deploy: create token with 21M supply
data:,{"p":"erc-20","op":"deploy",
  "tick":"eths","max":"21000000","lim":"1000"}

// Mint: claim 1000 units
data:,{"p":"erc-20","op":"mint",
  "tick":"eths","amt":"1000"}
\end{lstlisting}

\subsection{Transaction Role Patterns}
\label{sec:tx_roles}

Our analysis of sender--receiver relationships in \Uone\ reveals three dominant transaction role patterns, highlighting how Ethscription transactions differ from conventional value-transfer transactions on the Ethereum blockchain.

\begin{itemize}[leftmargin=*, nosep]
\item \textit{Self-send} ($70.7\%$ of \Uone): Sender and receiver are identical. This represents the canonical ``create'' behavior, where the inscriber records a data artifact to their own address. Rather than transferring value, these transactions primarily serve as on-chain publication events, highlighting the inscription-oriented nature of the workload.
\item \textit{Peer transfer} ($14.9\%$): Sender and receiver are distinct non-null addresses, corresponding to genuine asset or ownership movement. Compared with traditional Ethereum usage, this category forms a minority, indicating that most activity is not driven by economic exchange but by data creation.
\item \textit{Null-address send} ($14.4\%$): Transactions targeting the zero address (\texttt{0x000...000}), largely associated with Facet Protocol operations. These transactions function as protocol-specific signaling or state-registration mechanisms rather than user-to-user interaction.
\end{itemize}

\subsection{Semantic Boundary: On-chain Payload / Off-chain Indexer}
\label{sec:semantic_boundary}

An Ethscription transaction stores only the raw payload. Token balances, ownership, and validity are derived entirely by off-chain indexers that parse payloads according to protocol specifications. This creates a fundamental semantic boundary: the on-chain layer guarantees data availability and ordering, while the off-chain layer interprets meaning. This role separation leads to several practical implications:

\begin{itemize}[leftmargin=*, nosep]
\item \textit{Parser divergence:} Different indexers may interpret edge cases differently, leading to inconsistent state views.
\item \textit{Spec drift:} Protocol specifications can change, retroactively altering the interpretation of historical payloads.
\item \textit{No on-chain enforcement:} Invalid operations (minting beyond supply limits, transferring unknown tokens) are stored on-chain but rejected by compliant indexers.
\end{itemize}

\subsection{Non-JSON Ethscriptions (\Uone$\setminus$\Utwo) as Out-of-Scope}
\label{sec:non_json_scope}

Non-JSON Ethscriptions - images (PNG, SVG, JPEG, GIF, WebP), plain text, HTML, and Facet Protocol operations---collectively constitute $24.2\%$ of \Uone. These payloads lack the structured operation grammar that defines \Utwo and serve different purposes: media inscriptions embed cultural artifacts, while Facet operations encode a distinct smart-contract-like execution model. We report their composition shares (\Cref{tab:composition}) as context for the \Utwo scoping decision, but exclude them from all core analyses (temporal dynamics, protocol metrics, lifecycle funnel, concentration, and footprint). This exclusion is by design: our research questions target the JSON operational workload, not the broader media inscription phenomenon.

\begin{center}
\fbox{%
\begin{minipage}{0.9\linewidth}
\textbf{Takeaway:}
Ethscriptions are calldata-resident payloads identified by a \texttt{data:} URI prefix (\Uone). The dominant subset (75.8\%) consists of JSON-encoded token operations (\Utwo) conforming to a grammar with required fields \texttt{p} and \texttt{op}. Token semantics exist solely in off-chain indexers, not on-chain consensus.
\end{minipage}}
\end{center}

\section{Data and Methodology}
\label{sec:methodology}

\subsection{Dataset Hierarchy and Statistical Constraints}
\label{sec:hierarchy}

We emphasize that our goal is to measure Ethscriptions as a \textit{specific calldata-resident inscription workload}, rather than to characterize Ethereum calldata usage in general. To avoid conflating Ethscriptions with arbitrary calldata-heavy transactions (\eg contract invocations, ABI parameters, or rollup data availability payloads), we define nested datasets and report all statistics with an explicit scope label. Let \Uzero denote the universe of all Ethereum mainnet transactions within our observation window. We extract \Uone as the set of \textit{Ethscription candidates}, defined operationally as transactions whose calldata begins with the hexadecimal encoding of the string \texttt{data:} (\ie prefix \texttt{0x646174613a}). From \Uone, we further derive \Utwo, the \textit{Ethscription set}, consisting of transactions whose decoded \texttt{data:} payload is valid JSON and conforms to an operation grammar (\eg containing required keys such as \texttt{p} and \texttt{op}, with optional fields such as \texttt{tick} and \texttt{amt}, plus protocol-specific extensions). Unless stated otherwise, \textit{all workload dynamics, protocol, lifecycle, participation, and footprint analyses in this paper are computed on \Utwo}, while \Uone is used only to establish the composition boundary (\eg the share of JSON vs.\ non-JSON inscriptions). This design ensures that our findings describe the Ethscription ecosystem, particularly its JSON operational substrate, without making claims about general calldata behavior.

\begin{itemize}
    \item \textbf{\Uzero-Ethereum transaction universe.} All Ethereum mainnet transactions in our observation window (genesis through block 19,314,267, February 26, 2024). Mentioned only for high-level context.

  \item \textbf{\Uone-Ethscription candidate set.} Transactions whose calldata begins with the hexadecimal encoding of the string \texttt{data:} (\ie prefix \texttt{0x646174613a}). This is the operational definition of Ethscriptions used in this study. $|\Uone| = 6{,}269{,}364$.

  \item \textbf{\Utwo-Ethscription set (primary analysis set).} The subset of \Uone where the decoded payload: (i)~is valid JSON, and (ii)~matches an Ethscription operation grammar (schema constraints: required keys \texttt{p} and \texttt{op}, with optional fields \texttt{tick}, \texttt{amt}, plus protocol-specific extensions). $|\Utwo| = 4{,}749{,}698$ ($75.8\%$ of \Uone).
\end{itemize}


\subsection{Data Collection}
\label{sec:data_collection}

We query the full Ethereum mainnet transaction history from a local archive node~\cite{etherscan2024}, scanning all blocks from genesis through block $19{,}314{,}267$ (February 26, 2024). For each transaction whose raw input data begins with \texttt{0x646174613a}, we extract the following information: transaction hash, block number, sender address (\texttt{from}), receiver address (\texttt{to}), \textsc{ETH} value, block timestamp, raw calldata (hex), \textsc{UTF-8} decoded payload, and input data length in bytes.

\para{Observation Window} Our dataset spans March 2017 to February 2024. While the Ethscriptions protocol was formally introduced in June 2023~\cite{lehman2023ethscriptions}, isolated \texttt{data:}-prefixed transactions appeared as early as March 2017 (11~precursor transactions). The ecosystem became active between June 2023 and February 2024 (9 months), during which $99.99\%$ of all \Uone transactions occurred.

\subsection{Dataset Summary}
\label{sec:dataset_summary}

\Cref{tab:dataset_u1} and \Cref{tab:dataset_u2} provide scope-aware summaries.

\begin{table}[tbh]
\centering
\caption{Dataset Summary: \Uone (Ethscription Candidates)}
\label{tab:dataset_u1}
\small
\begin{tabular}{l|r}
\toprule
\multicolumn{1}{c}{\textbf{Metric}} & \multicolumn{1}{c}{\textbf{Value}} \\
\midrule
\rowcolor{gray!10}
Total $|\Uone|$ & 6,269,364 \\
\midrule
Active period & Jun 2023 -- Feb 2024 \\
Block range & 9,046,974 -- 19,314,267 \\
Unique senders & 212,039 \\
Unique receivers & 176,102 \\
Self-send (\texttt{from}=\texttt{to}) & 4,435,375 (70.7\%) \\
Null-address sends & 903,621 (14.4\%) \\
Zero-value transactions & 6,130,885 (97.8\%) \\
\midrule
\rowcolor{gray!10}
\multicolumn{2}{l}{\textit{Payload size (bytes)}} \\
\midrule
\quad Median / Mean & 75 / 238 \\
\quad P95 / P99 / Max & 419 / 2,870 / 130,935 \\
\bottomrule
\end{tabular}
\end{table}

\begin{table}[tbh]
\centering
\caption{Dataset Summary: \Utwo (Ethscription Operations)}
\label{tab:dataset_u2}
\small
\begin{tabular}{l|r}
\toprule
\multicolumn{1}{c}{\textbf{Metric}} & \multicolumn{1}{c}{\textbf{Value}} \\
\midrule
\rowcolor{gray!10}
Total $|\Utwo|$ & 4,749,698 \\
\midrule
\rowcolor{gray!10}
Share of \Uone & 75.8\% \\
Unique senders & $\leq$212,039\textsuperscript{$\dagger$} \\
Unique receivers & $\leq$176,102\textsuperscript{$\dagger$} \\
Distinct protocol identifiers (\texttt{p}) & 30+ \\
Distinct operation types (\texttt{op}) & 20+ \\
Distinct token tickers (\texttt{tick}) & 1,000+ \\
\midrule
\rowcolor{gray!10}
\multicolumn{2}{l}{\textit{Operation breakdown (\Utwo)}} \\
\midrule
\quad mint & 4,461,324 (93.9\%) \\
\quad deploy & 22,172 (0.5\%) \\
\quad transfer / proxy\_transfer & 77,508 (1.6\%) \\
\quad trade (freeze\_sell, list, buy, sell) & 76,752 (1.6\%) \\
\quad other (stake, claim, etc.) & 111,942 (2.4\%) \\
\midrule
\rowcolor{gray!10}
\multicolumn{2}{l}{\textit{Payload size (bytes, JSON only)}} \\
\midrule
\quad Median & $\sim$75 \\
\quad P95 & $\sim$120 \\
\bottomrule
\multicolumn{2}{l}{\scriptsize\textsuperscript{$\dagger$}Upper bound: subset of \Uone addresses; exact \Utwo count is a subset.}
\end{tabular}
\end{table}

\subsection{Processing Pipeline}
\label{sec:pipeline}

To obtain a scope-consistent Ethscription workload, we construct a deterministic multi-stage extraction pipeline that transforms raw Ethereum transactions into the analytical dataset used throughout this study. The pipeline separates identification, validation, and analysis, ensuring that inscription activity is isolated from general calldata usage.

\subsubsection{Pipeline Definition}
Our analysis proceeds through four sequential stages (\Cref{fig:methodology}):

\begin{itemize}[leftmargin=*, nosep]

\item \textit{Stage 0: Scan \Uzero.} 
All Ethereum mainnet transactions are scanned for inputs with prefix \texttt{0x646174613a} (the hexadecimal encoding of the \texttt{data:} URI scheme).

\item \textit{Stage1: Prefix Filter $\rightarrow$ \Uone.} Transactions matching the prefix are validated for structural consistency (block range, timestamp integrity, and address format), producing the Ethscription candidate set~\Uone.

\item \textit{Stage2: Decode and MIME Classification.} Payloads are decoded and classified by content type to establish the composition boundary of the workload, distinguishing JSON operational inscriptions from media or free-form payloads. No analytical metrics are computed at this stage.

\item \textit{Stage3: JSON Validation and Grammar Parsing $\rightarrow$ \Utwo.} JSON payloads are validated against the operation grammar defined in \S\ref{sec:operation_model}. Transactions satisfying required fields (\texttt{p}, \texttt{op}) form the primary analysis set~\Utwo.

\item \textit{Stage4: Analytics (on \Utwo).} All subsequent measurements, including temporal dynamics, protocol metrics, lifecycle funnel behavior, participation concentration, and persistent footprint studies, are conducted primarily on~\Utwo.
\end{itemize}

\subsubsection{Dataset Reduction Funnel}
The staged pipeline forms a measurable filtering funnel, summarized in \Cref{tab:funnel_details}. 
Starting from the candidate set~\Uone, we remove non-JSON payloads to isolate operational inscriptions, producing \Utwo, which accounts for 75.8\% of all detected inscriptions.

\begin{table}[tbh]
\centering
\caption{Filtering funnel from \Uone\ to \Utwo.}
\label{tab:funnel_details}
\small
\begin{tabular}{l|r|r}
\toprule
\multicolumn{1}{c|}{\textbf{Stage}} &
\multicolumn{1}{c|}{\textbf{Count}} &
\multicolumn{1}{c}{\textbf{\% of Prior}} \\
\midrule
\rowcolor{gray!10}
\Uzero (all Ethereum txs) & --- & --- \\
\rowcolor{gray!10}
\Uone (data: prefix) & 6,269,364 & --- \\
\quad Non-JSON excluded & 1,519,666 & 24.2\% \\
\quad \quad Images & 274,314 & 4.4\% \\
\quad \quad Short text & 683,115 & 10.9\% \\
\quad \quad Facet Protocol & 449,909 & 7.2\% \\
\quad \quad Long text & 78,911 & 1.3\% \\
\quad \quad HTML & 25,768 & 0.4\% \\
\quad \quad Other & 7,649 & 0.1\% \\
\rowcolor{gray!10}
\Utwo (JSON + grammar) & 4,749,698 & 75.8\% \\
\bottomrule
\end{tabular}
\end{table}

\subsubsection{Identification Validity.}
False positives are expected to be negligible because the prefix \texttt{0x646174613a} is highly specific and unlikely to collide with standard ABI encoding patterns. False negatives may arise from non-standard encodings that omit the canonical prefix; however, without external ground truth, such cases cannot be reliably enumerated. Our definition therefore prioritizes alignment with the canonical Ethscription specification rather than heuristic reconstruction.






\subsubsection{Robustness and Sensitivity Analysis}

In the following, we evaluate whether the analytical results depend on parsing assumptions or grammar strictness.

\para{Grammar Strictness}
Relaxing the grammar from requiring both \texttt{p} and \texttt{op} to requiring only \texttt{p} increases \Utwo\ by $<$0.1\%, while requiring only \texttt{op} produces a $<$0.2\% increase. This indicates that nearly all JSON payloads already conform to the full grammar.

\para{Alternative Parsing}
Using strict JSON parsing (rejecting trailing commas or single-quoted strings) versus lenient parsing changes \Utwo\ by $<$0.05\%, suggesting that most payloads are well-formed and likely machine-generated.

\para{Content-Type Classification}
Our MIME classification relies on \texttt{data:} URI headers. For shorthand payloads (\texttt{data:,\{...\}}) without an explicit media type, classification falls back to JSON parsing. Explicit \texttt{data:application/json} headers are rare ($<$1\% of \Utwo), indicating that most JSON payloads rely on implicit interpretation rather than explicit MIME annotation.

\subsection{Ethscription Grammar and Normalization}
\label{sec:grammar}

We define the Ethscription grammar as follows, based on observed transaction payloads:

\begin{lstlisting}[caption={Ethscription grammar (core fields).}, label=lst:grammar, captionpos=t]
{
  "p":    <string>,  // REQUIRED: protocol id
  "op":   <string>,  // REQUIRED: operation type
  "tick": <string>,  // OPTIONAL: token ticker
  "amt":  <string|number>,  // OPTIONAL: amount
  ...                // protocol-specific extensions
}
\end{lstlisting}

\para{Required Fields} \texttt{p} (protocol identifier, \eg \texttt{erc-20}, \texttt{ierc-20}) and \texttt{op} (operation type, \eg \texttt{deploy}, \texttt{mint}, \texttt{transfer}).

\para{Optional Fields} \texttt{tick} (token ticker symbol), \texttt{amt} (amount, encoded as string or integer), plus protocol-specific extensions (\eg \texttt{lim}, \texttt{max} for deploy operations; \texttt{pow} for proof-of-work parameters in \texttt{ierc-pow}).

\para{Normalization}
We canonicalize operation strings (e.g., merging aliases such as \texttt{Transfer}/\texttt{transfer}), unify numeric representations (e.g., \texttt{amt} as string vs.\ integer), and normalize whitespace and encoding.

\subsection{Metrics and Analysis Methods}
\label{sec:metrics}

Below metrics are computed on \Utwo unless explicitly noted.

\begin{itemize}
    \item \textit{Temporal Dynamics.} We measure daily and monthly transaction volumes and segment activity into bootstrap, expansion, and saturation phases using reproducible criteria derived from growth and stabilization patterns.

    \item \textit{Protocol Metrics.} We track the number of distinct \texttt{p} identifiers, their entry and exit timing, and their market-share trajectories throughout the active observation period to characterize protocol-level competition and turnover.

    \item \textit{Lifecycle Funnel.} 
    Based on the lifecycle conversion pipeline (deploy $\rightarrow$ mint $\rightarrow$ transfer $\rightarrow$ trade), we quantify stage transitions, including the deploy-to-mint amplification and mint-to-transfer drop-off rates, to characterize participation dynamics.

    \item \textit{Concentration.} We evaluate ecosystem concentration using the Gini coefficient over sender transaction-count distributions, complemented by Lorenz curves and Top-N concentration ratios. Sender and receiver concentration are computed on \Uone (all Ethscriptions) to capture structural ecosystem properties, while \Utwo-specific sender concentration is additionally reported where relevant.

    \item \textit{Persistent Footprint.} We quantify the total calldata bytes attributable to \Utwo and measure its proportional footprint within the broader \Uone calldata usage.
\end{itemize}




\section{Measurement Results}
\label{sec:results}
We present our measurement results in this section. Specifically, we report only aggregate data statistics and do not link addresses to real-world identities.

All measurements in this section are computed on \Utwo (Ethscription, $n = 4{,}749{,}698$) unless explicitly noted otherwise. Section~\Cref{sec:results} is measurement-only; interpretation and ``why'' analysis are deferred to \Cref{sec:discussion}.

\subsection{Scope Establishment: Composition and Filtering Results}
\label{sec:scope_establishment}

We first quantify the scope of our measured workload to prevent conflating Ethscriptions with general calldata usage. Starting from the Ethereum transaction universe \Uzero, we extract the Ethscription candidate set \Uone by identifying transactions whose calldata begins with the \texttt{data:} URI prefix. We then classify decoded \texttt{data:} payloads by content type (\Cref{tab:composition}) and observe that a substantial fraction of \Uone consists of JSON-formatted payloads. Since the operational ecosystem of interest is driven by structured operations, we further restrict to the Ethscription set \Utwo, retaining only those \Uone transactions whose payload is valid JSON and matches our operation grammar (\eg required keys \texttt{p} and \texttt{op}, with optional fields such as \texttt{tick} and \texttt{amt}). We report \textbf{(i)}~the composition ratio $|\Utwo|/|\Uone|$, establishing the dominance of JSON operations within Ethscriptions, and \textbf{(ii)}~a breakdown of exclusion reasons when constructing \Utwo from \Uone (non-JSON media inscriptions; malformed or undecodable payloads; JSON that fails grammar constraints). 

\para{Composition Ratio} $|\Utwo|/|\Uone| = 4{,}749{,}698 / 6{,}269{,}364 = 75.8\%$, confirming JSON operational dominance.

\para{Exclusion Breakdown (\Uone$\rightarrow$\Utwo)}
We exclude payloads that fail grammar constraints or are non-JSON. 
Non-JSON payloads dominate the exclusions (24.2\% of \Uone), including short text (10.9\%), Facet Protocol (7.2\%), images (4.4\%), long text (1.3\%), HTML (0.4\%), and others (0.1\%). 
Malformed/undecodable payloads and JSON missing required fields (e.g., \texttt{p}/\texttt{op}) each account for $<$0.1\%. 

The resulting \Utwo retains 75.8\% of \Uone.

\subsection{RQ1 --- Workload Dynamics of Ethscription (\Utwo)}
\label{sec:rq1}

\subsubsection{{Longitudinal Timeline (\Utwo)}}

\begin{figure*}[htb]
    \centering
    \subfigure[Full timeline (log scale). 99.99\% of \Uone occurs in the 9-month active period.]{
    \begin{minipage}[t]{0.45\textwidth}
    \centering
    \includegraphics[width=\linewidth]{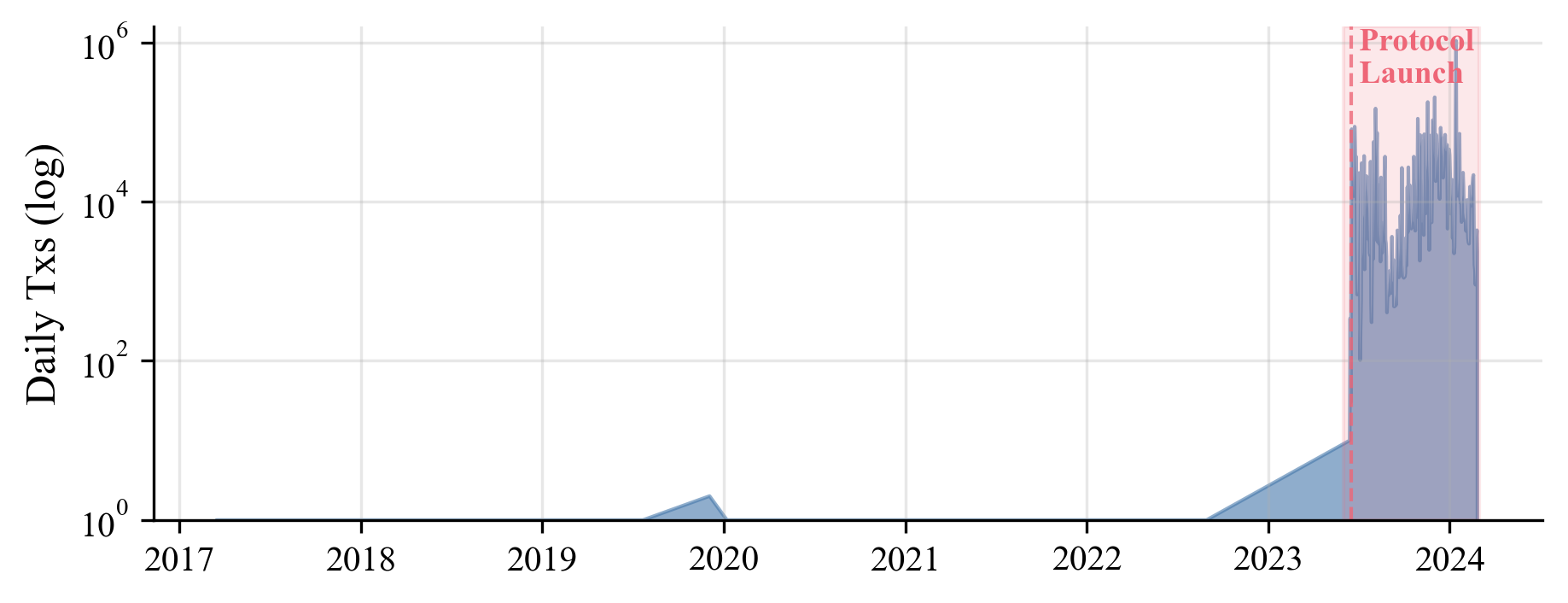}
    \end{minipage}
    \label{fig:daily_full}
    }
    \subfigure[Active period: self-sends (minting) dominate throughout.]{
    \begin{minipage}[t]{0.45\textwidth}
    \centering
    \includegraphics[width=\linewidth]{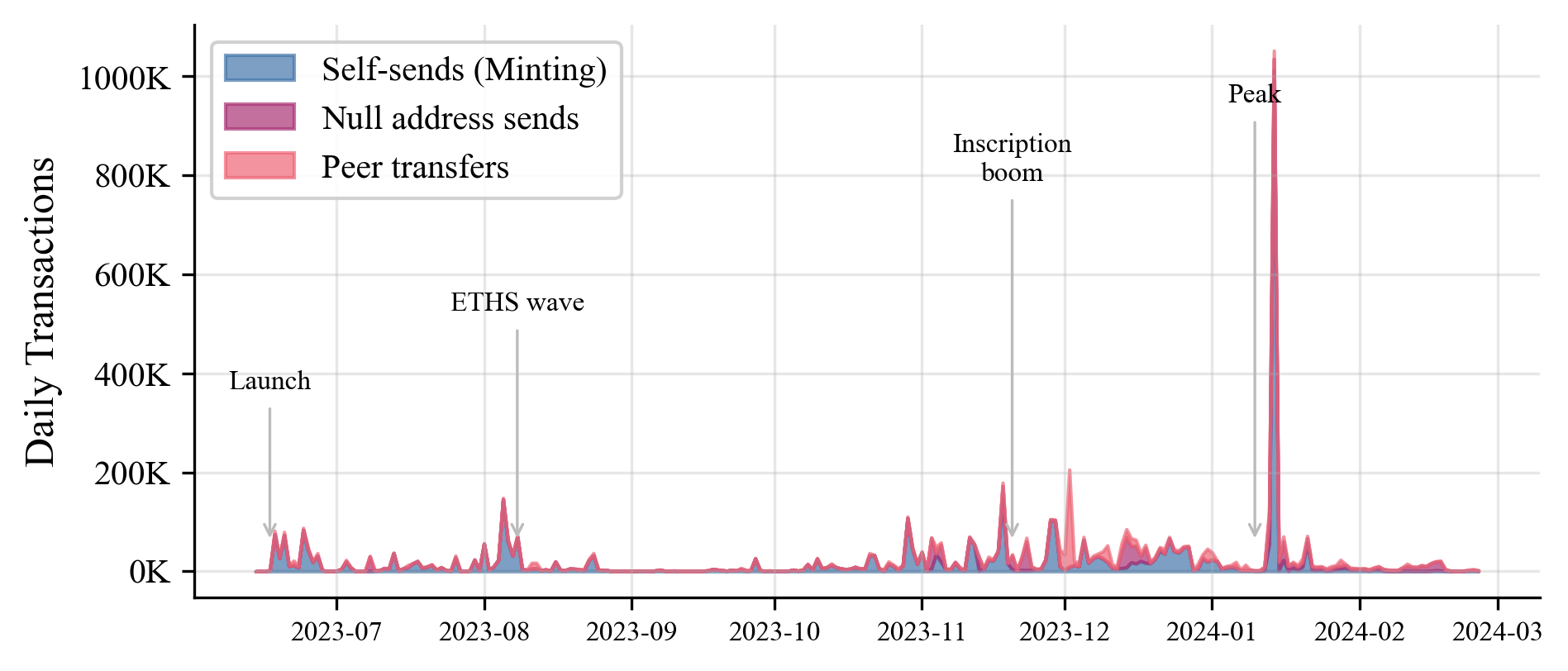}
    \end{minipage}
    \label{fig:daily_zoom}
    }
    \caption{Transaction volume of Ethscriptions. (a) Full timeline (log scale) showing that 99.99\% of \Uone occurs within a short active period. (b) Active-period view (Jun 2023–Feb 2024), highlighting three phases: bootstrap, expansion, and saturation/decline.}
    \label{fig:daily_activity}
\end{figure*}

\Cref{fig:daily_activity} shows the overall transaction volume of \Utwo, while \Cref{fig:daily_zoom} provides a zoomed-in view of the active period, where a clear three-phase lifecycle becomes apparent:

\begin{itemize}[leftmargin=*, nosep]
\item \textit{Phase-1: Bootstrap} (Jun--Jul 2023). The protocol launches on June 17, 2023. JSON operations emerge immediately with image and text inscriptions. The \Utwo volume is moderate, primarily driven by early-adopter experimentation.

\item \textit{Phase-2: Expansion} (Aug--Oct 2023). JSON operations rapidly come to dominate the ecosystem. \Utwo monthly volume peaks at $\sim$$434$K in August as the \texttt{erc-20} protocol triggers large-scale minting activity, and subsequently oscillates as successive protocol variants enter and compete.

\item \textit{Phase-3: Saturation \& Decline} (Nov 2023--Feb 2024).  Monthly \Utwo volume continues to escalate, exceeding $900$K in November and reaching a peak in January 2024, followed by a sharp decline in February ($<$$200$K for the partial month), indicating rapid cooling and workload saturation.
\end{itemize}

\subsubsection{{Payload Size Features \& Temporal Evolution (\Utwo)}}
JSON payloads are compact, with a median size of $75$~bytes. 
This distribution is tightly concentrated because the dominant operation (mint) uses a fixed-format payload (\texttt{p}, \texttt{op}, \texttt{tick}, \texttt{amt}) with minimal variation. 
Deploy operations are slightly larger (median $\sim$$120$~bytes) due to additional fields such as \texttt{max} and \texttt{lim}. 
By contrast, while \Uone\ exhibits a heavy-tailed distribution (driven by image inscriptions up to $131$~KB), \Utwo\ payload sizes remain highly homogeneous.

\para{Temporal Evolution} Payload sizes remain stable across the active period. The median stays within $70$--$80$~bytes throughout all three phases, reflecting the fixed structure of mint operations that dominate \Utwo. A slight increase in mean payload size during Phase~3 (Nov 2023--Feb 2024) corresponds to the emergence of \texttt{ierc-pow}, whose proof-of-work parameters (\texttt{pow}, \texttt{nonce}) add $\sim$$20$--$40$ bytes per payload. Overall, payload size is driven by operation type rather than temporal factors.

\begin{center}
\fbox{%
\begin{minipage}{0.9\linewidth}
\textbf{Takeaway:}
The Ethscription workload compresses a full boom-bust cycle into $9$~months, transitioning from bootstrap through expansion to saturation. Compact, homogeneous payloads (median 75~bytes) make individual operations negligibly cheap, enabling high-volume speculative minting that drives the rapid phase transitions.
\end{minipage}}
\end{center}

\subsection{RQ2 --- Protocol Landscape and Lifecycle (\Utwo)}
\label{sec:rq2}

\subsubsection{{Protocol Identifier Distribution}}
We identify $30$+ distinct protocol identifiers (\texttt{p} field values) in \Utwo. \Cref{tab:protocols} presents the top $15$ identifiers by transaction volume.

\begin{table}[t]
\centering
\caption{Top 15 Ethscription Protocols by Volume (\Utwo)}
\label{tab:protocols}
\small
\begin{tabular}{c|r|r|p{2.8cm}}
\toprule
\textbf{Protocol (\texttt{p})} & \multicolumn{1}{c|}{\textbf{Txs}} & \textbf{\%} & \textbf{Description} \\
\midrule
erc-20 & 3,069,682 & 64.6 & Base standard \\
\rowcolor{gray!10}
ierc-20 & 677,814 & 14.3 & ``Improved'' variant \\
ierc-pow & 321,533 & 6.8 & PoW minting \\
\rowcolor{gray!10}
rerc-20 & 278,744 & 5.9 & ``Recursive'' variant \\
terc-20 & 82,780 & 1.7 & ``Typed'' variant \\
\rowcolor{gray!10}
esc-20 & 69,774 & 1.5 & Ethscriptions-native \\
erc-cash & 45,892 & 1.0 & Cash-like transfers \\
\rowcolor{gray!10}
brc-20 & 23,416 & 0.5 & BRC-20 naming port \\
asc-20 & 18,821 & 0.4 & Arbitrum name port \\
\rowcolor{gray!10}
oft-20 & 12,703 & 0.3 & ``Open Fair Token'' \\
ebrc-20 & 8,128 & 0.2 & ETH-BRC hybrid \\
\rowcolor{gray!10}
grc-20 & 4,333 & 0.1 & Generic variant \\
bsc-20 & 4,297 & 0.1 & BSC naming port \\
\rowcolor{gray!10}
src-20 & 4,088 & 0.1 & Stacks naming port \\
zrc-20 & 3,716 & 0.1 & zkSync naming port \\
\bottomrule
\end{tabular}
\end{table}

\para{Naming Convention} While 30+ protocol identifiers exist on the Ethereum mainnet, we find that names such as \texttt{brc-20}, \texttt{asc-20}, \texttt{bsc-20} do \textit{not} necessarily represent cross-chain activity; users on Ethereum created protocols borrowing names from other chain ecosystems. The dominant pattern is \texttt{X-20} where \texttt{X} varies by chain prefix.

\subsubsection{Protocol Dynamics}
\label{sec:proto_dynamics}

\begin{figure*}[htb]
    \centering
    \subfigure[Monthly volume by protocol showing entry and exit dynamics.]{
    \begin{minipage}[t]{0.45\textwidth}
    \centering
    \includegraphics[width=\linewidth]{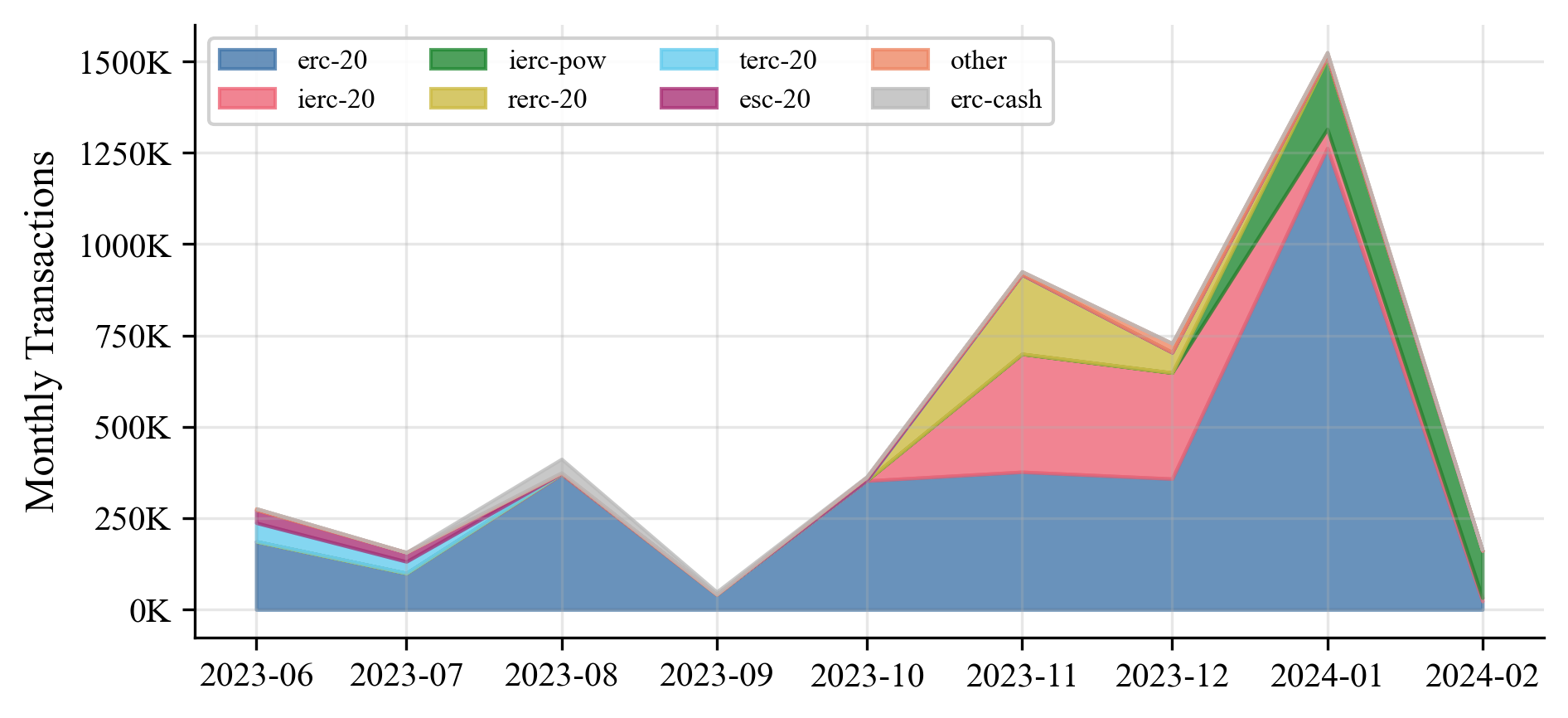}
    \end{minipage}
    \label{fig:proto_vol}
    }
    \subfigure[Market-share trajectories of top 8 protocols.]{
    \begin{minipage}[t]{0.45\textwidth}
    \centering
    \includegraphics[width=\linewidth]{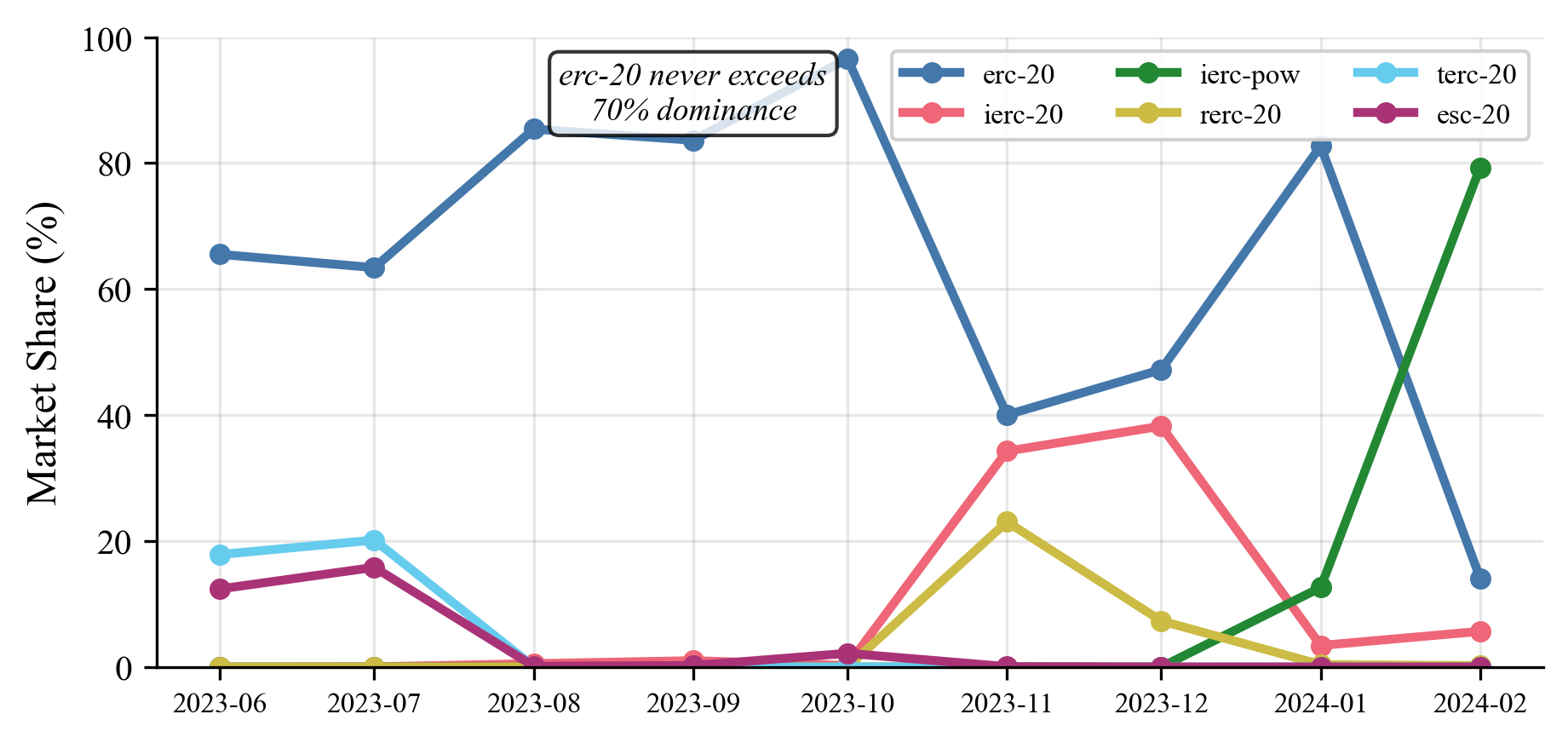}
    \end{minipage}
    \label{fig:proto_share}
    }
    \caption{Protocol adoption dynamics in \Utwo. 
    (a) Monthly transaction volume by protocol, showing rapid entry and exit of competing protocols. 
    (b) Market-share trajectories of the top protocols, illustrating short-lived dominance and frequent turnover.}
    \label{fig:protocol_evolution}
\end{figure*}

\Cref{fig:protocol_evolution} shows monthly protocol adoption dynamics.

\begin{itemize}[leftmargin=*, nosep]
\item \texttt{erc-20} remains the baseline protocol throughout the period, but its dominance is limited, never exceeding $77$\% of monthly share (October 2023 peak).
\item New protocols can rapidly gain traction upon introduction. For instance, \texttt{ierc-20} emerges in November 2023 and immediately captures $28$\% of activity.
\item This pattern repeats with later entrants: \texttt{ierc-pow} surges in January--February 2024, driven by PoW minting.
\item Overall, these dynamics indicate a highly fluid ecosystem in which protocols can rise quickly but rarely sustain dominance, reflecting the low barrier to protocol creation.
\end{itemize}

Although a small number of protocols account for the majority of activity,  the ecosystem still exhibits a long-tail structure (Appendix~\ref{app:longtail}).
The long-tail distribution includes more than 15 additional protocol identifiers (each $<$$4$K transactions), reinforcing persistent fragmentation.
No single protocol achieves network-effect dominance.

\para{Protocol Churn} Over the $9$-month active period, new protocol identifiers enter at a rate of about 3--4 per month. We define a protocol as \emph{active} in a month if it appears in at least
{[T]} transactions during that month (we use $T=100$ by default to
filter out one-off spam identifiers). Of the $30$+ observed protocols, fewer than $10$ sustain activity beyond their debut month. The monthly protocol churn rate (fraction of active protocols that are new entrants) ranges from $20$\% to $50$\%, confirming that protocol creation is low-cost and experimentation is persistent. No protocol that entered after \texttt{erc-20} (the founding standard) has displaced it from plurality position, yet \texttt{erc-20}'s monthly share fluctuates between $45$\% and $77$\%, indicating ongoing competitive pressure from successive entrants.

\subsubsection{{Operation Type Distribution (\Utwo)}}

\Cref{tab:operations} reports the operation lifecycle distribution.

\subsubsection{{Lifecycle Funnel and Conversion Efficiency}}

The lifecycle funnel reveals dramatic drop-off:
\begin{equation}
\text{deploy} \xrightarrow{201\times} \text{mint} \xrightarrow{57.6:1} \text{transfer} \xrightarrow{\sim 1:1} \text{trade}
\label{eq:lifecycle}
\end{equation}
\begin{itemize}[leftmargin=*, nosep]
\item \textbf{Deploy $\rightarrow$ Mint amplification:} 22,172 deploys produce 4,461,324 mints ($201\times$). Each token deployment triggers extensive minting.
\item \textbf{Mint $\rightarrow$ Transfer drop-off:} 57.6:1 ratio. The vast majority of minted tokens are never transferred.
\item \textbf{Transfer $\rightarrow$ Trade:} $\sim$1:1 ratio. Among tokens that move, roughly equal numbers enter marketplace operations.
\end{itemize}

\subsubsection{{Token Ticker Surface (\Utwo)}}

\Cref{tab:top_tokens} presents the top tokens. The leading token \texttt{nodes} accounts for 22.8\% of \Utwo. Token names span functional references (\texttt{rETH}, \texttt{powETH}, \texttt{etfDay}) and meme-culture signaling (\texttt{pepe}, \texttt{dumb}, \texttt{Troll}).

\begin{center}
\fbox{%
\begin{minipage}{0.9\linewidth}
\textbf{Takeaway:}
Zero-cost protocol creation leads to persistent fragmentation: over 30 competing \texttt{p}-identifiers emerge without convergence. The lifecycle funnel (201$\times$ deploy-to-mint amplification; 57.6:1 mint-to-transfer drop-off) further shows that most minted tokens are never transferred or traded, indicating predominantly speculative minting.
\end{minipage}}
\end{center}

\subsection{RQ3 --- Participation Structure and Footprint}
\label{sec:rq3}

\subsubsection{{Sender/Receiver Concentration}}

\begin{figure}[!t]
    \centering
    \subfigure[Lorenz curve for sender activity (Gini = 0.858, \Uone), zoomed to top 5\%.]{
    \begin{minipage}[t]{0.4\textwidth}
    \centering
    \includegraphics[width=\linewidth]{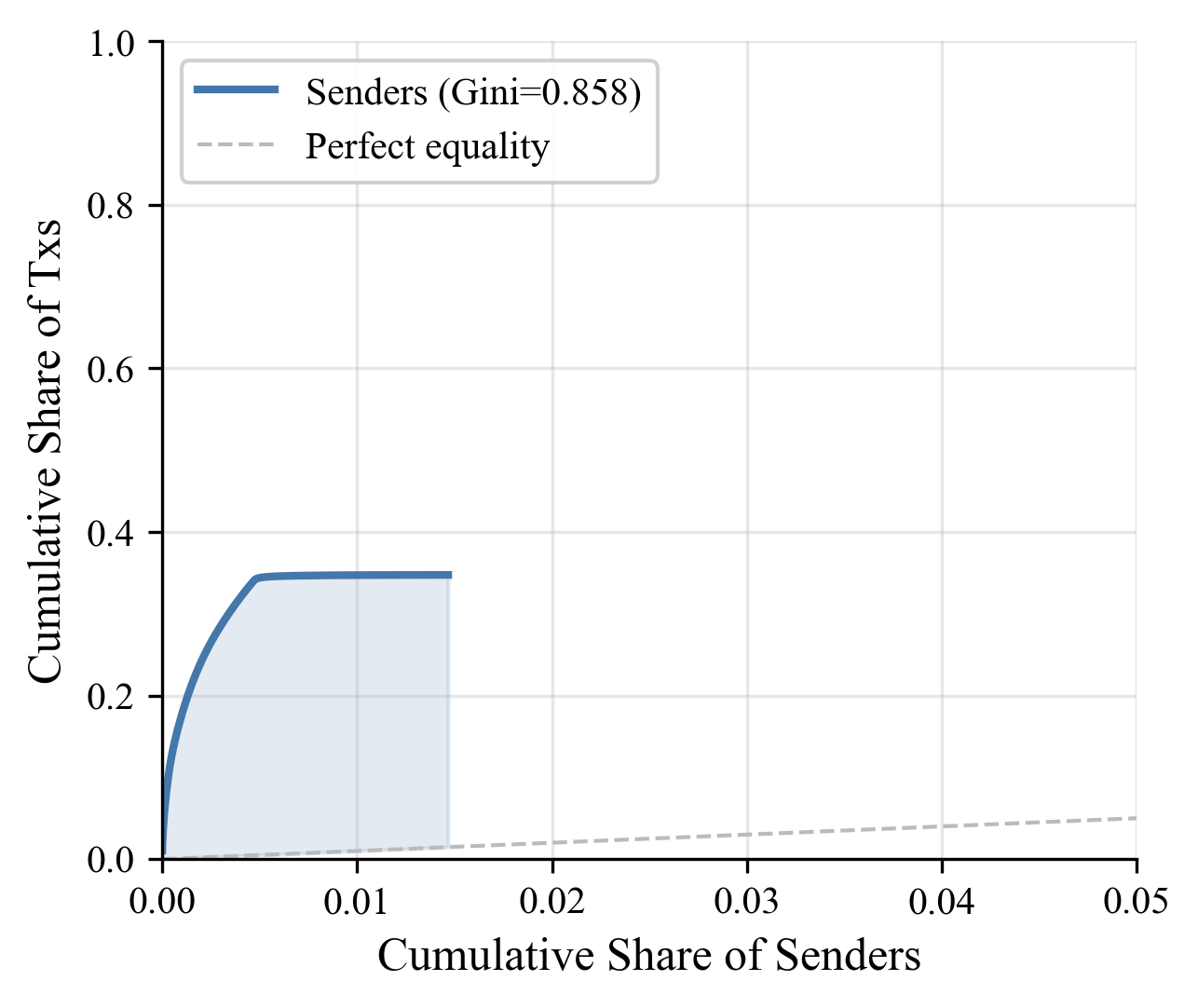}
    \end{minipage}
    \label{fig:lorenz}
    }
    \subfigure[Top-N concentration: top 1K senders (0.5\%) produce 34.1\% of transactions (\Uone).]{
    \begin{minipage}[t]{0.4\textwidth}
    \centering
    \includegraphics[width=\linewidth]{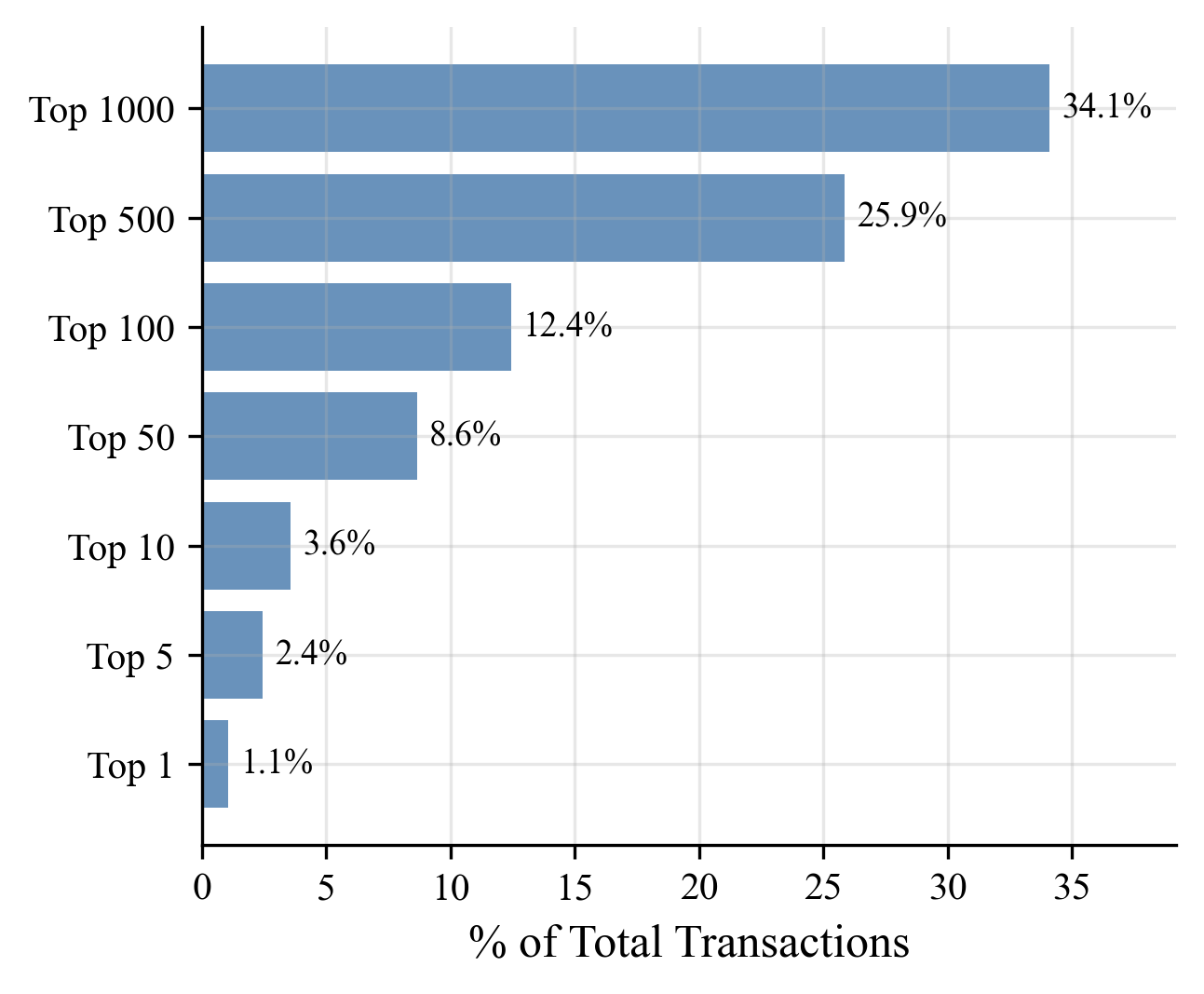}
    \end{minipage}
    \label{fig:topn}
    }
    \caption{Address concentration in \Uone. 
(a) Lorenz curve showing strong inequality in sender activity (Gini = 0.858). 
(b) Top-$N$ concentration, illustrating that a small fraction of addresses generates a large share of transactions.}
    \label{fig:network}
\end{figure}

The ecosystem exhibits extreme address concentration. \Cref{fig:network} presents Lorenz curves and Top-N analysis.

\para{Sender Concentration (\Uone)} The Gini coefficient for sender activity across all Ethscriptions is $0.858$. The top $1{,}000$ senders ($0.5$\% of $212{,}039$ unique senders) account for $34.1$\% of all \Uone transactions. In addition, the single most active sender generated $66,130$ transactions ($1.06$\%), indicating a highly concentrated participation structure.

\para{Receiver Concentration (\Uone)} The Gini coefficient for receiver activity is $0.900$. The null address alone received $903{,}621$ transactions ($14.4$\% of \Uone), and the Facet Protocol contract address received $445{,}317$ ($7.1$\%).

\para{Note on Scope} Concentration metrics are computed on \Uone because address activity spans both JSON and non-JSON inscriptions, and the structural inequality characterizes participation in the broader Ethscription ecosystem. Since \Utwo constitutes $75.8$\% of \Uone and JSON minting is the dominant operation, the sender concentration within \Utwo is expected to be comparable or higher (mint-heavy users dominate both sets). We flag this explicitly as a scope caveat (\Cref{sec:limitations}).

\begin{figure}[htb]
    \centering
    \subfigure[Operation distribution: \texttt{mint} dominates at 93.9\% of \Utwo.]{
    \begin{minipage}[t]{0.44\textwidth}
    \centering
    \includegraphics[width=\linewidth]{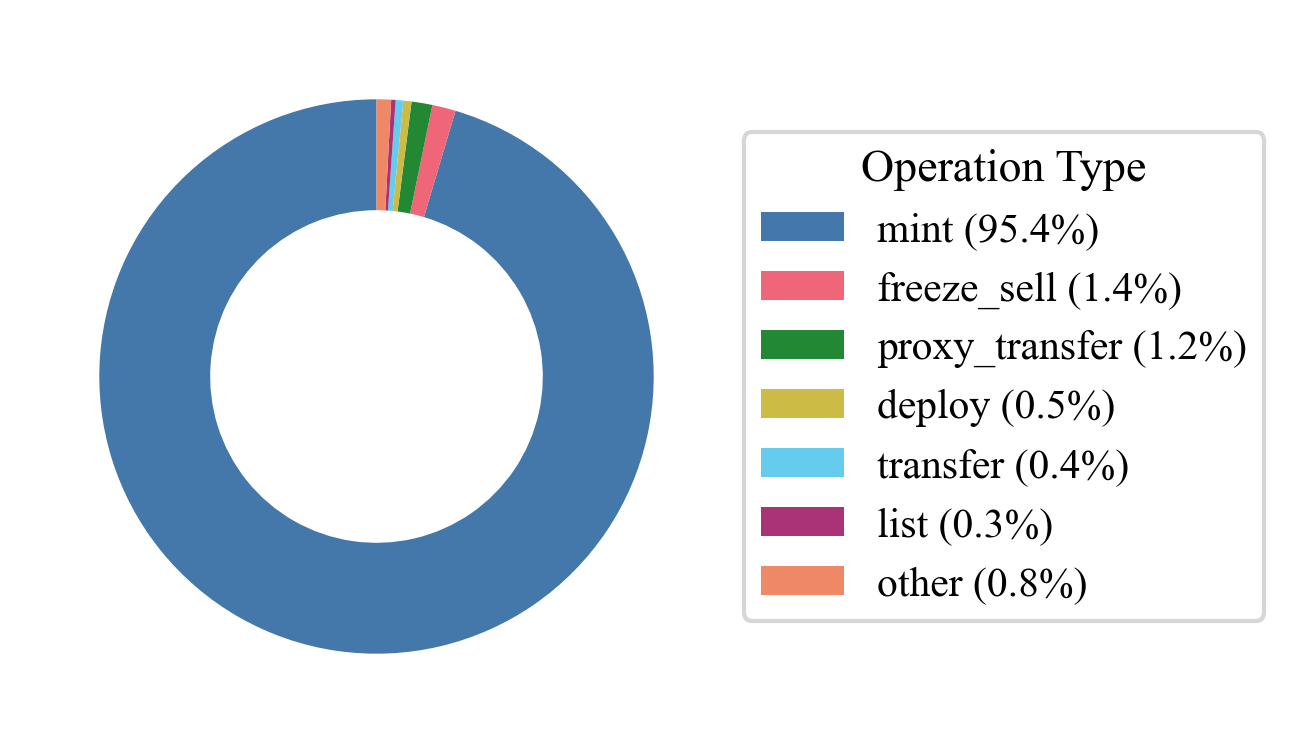}
    \end{minipage}
    \label{fig:ops_donut}
    }
    \subfigure[Top 15 token tickers by transaction volume (\Utwo).]{
    \begin{minipage}[t]{0.4\textwidth}
    \centering
    \includegraphics[width=\linewidth]{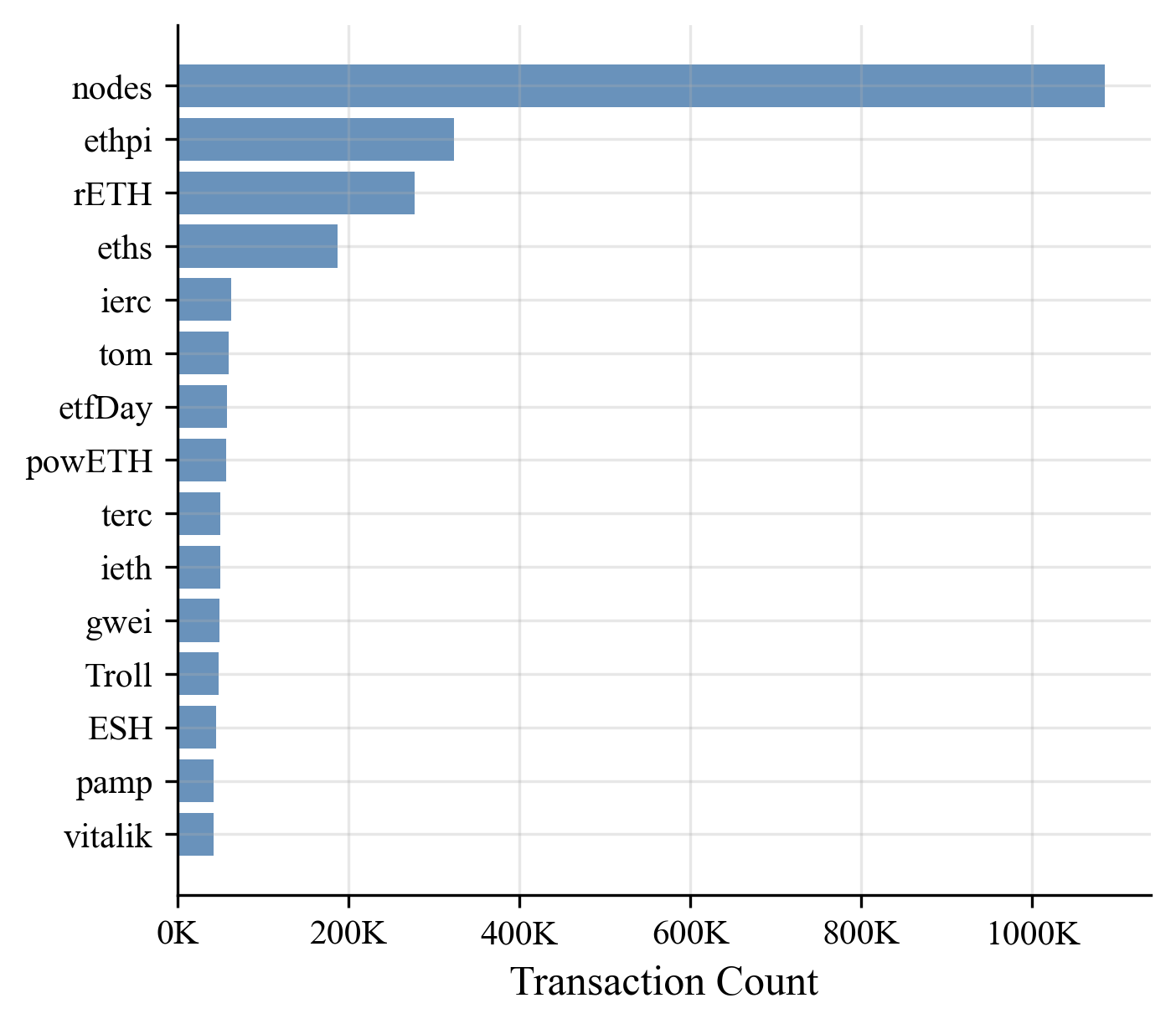}
    \end{minipage}
    \label{fig:top_tokens_bar}
    }
    \caption{Operation and token analysis (\Utwo). The funnel narrows dramatically: 22K deploys $\rightarrow$ 4.46M mints $\rightarrow$ 78K transfers $\rightarrow$ 77K trades.}
    \label{fig:operations}
\end{figure}

The power-law distribution indicates that a small fraction of addresses generates a disproportionate share of both transactions and the resulting storage externality.

\begin{table}[b]
\centering
\caption{Operations: Volumes and Funnel Ratios (\Utwo)}
\label{tab:operations}
\small
\begin{tabular}{c|l|r|r}
\toprule
\textbf{Stage} & \multicolumn{1}{c|}{\textbf{Operations}} & \textbf{Count} & \textbf{Share} \\
\midrule
Deploy & deploy & 22,172 & 0.5\% \\
\rowcolor{gray!10}
Mint & mint & 4,461,324 & 93.9\% \\
Transfer & transfer, proxy\_transfer & 77,508 & 1.6\% \\
\rowcolor{gray!10}
Trade & freeze\_sell, list, buy, sell & 76,752 & 1.6\% \\
Other & stake, claim, pack, etc. & 111,942 & 2.4\% \\
\bottomrule
\end{tabular}
\end{table}

\subsubsection{{Concentration within \Utwo (JSON-only)}}
\label{sec:u2_concentration}

To match our scope discipline, we also compute concentration on the primary
analysis set \Utwo. Let $c(a)$ denote the number of \Utwo transactions sent by
address $a$. We compute the sender Gini coefficient over the multiset
$\{c(a)\}$ as well as Top-$N$ shares.

\para{Sender Concentration (\Utwo)} 
The sender Gini coefficient within \Utwo is expected to be comparable to or slightly higher than the U1 value (Gini = 0.858), since JSON minting dominates activity and heavy-volume inscribers appear in both sets.

The top 1,000 senders account for 34.1\% of all Ethscription transactions in U1, and the dominance of mint-heavy activity suggests a comparable within U2.

\para{Scope Note} We retain \Uone-level receiver concentration for the null
address and Facet-related endpoints as a characterization of the broader
Ethscription ecosystem, but \Utwo-level results provide a scope-aligned
view for JSON operational substrates.

\subsubsection{{Persistent Data Footprint}} We next examine the persistent storage externality introduced by Ethscriptions, quantifying the calldata footprint added to Ethereum's chain history.

\para{Footprint Definition} We distinguish two byte measures:
(i)~\textit{input bytes} $B_{\mathrm{in}}$, the total number of bytes in the
transaction input field (\texttt{tx.input}), which includes the \texttt{data:}
prefix and any media-type headers; and (ii)~\textit{payload bytes}
$B_{\mathrm{payload}}$, the number of bytes after decoding the \texttt{data:}
URI payload (i.e., excluding the prefix/header). Unless stated otherwise, we
report footprint in terms of $B_{\mathrm{in}}$ because it is the on-chain data
replicated by all full nodes.

\para{Exact vs.\ Estimated Footprint} For \Uone, we compute the exact footprint
as $\sum_{\mathrm{tx}\in\Uone} B_{\mathrm{in}}(\mathrm{tx})$. For \Utwo, we
report both: (a)~the exact footprint $\sum_{\mathrm{tx}\in\Utwo}
B_{\mathrm{in}}(\mathrm{tx})$, and (b)~a median-based back-of-the-envelope
estimate $|\Utwo| \times \widetilde{B}_{\mathrm{payload}}$ to provide intuition
about per-operation compactness. We explicitly label any median-based numbers
as estimates because \Uone exhibits a heavy-tailed size distribution due to
media inscriptions.

\begin{table}[t]
\centering
\caption{Top 10 Token Tickers by Volume (\Utwo)}
\label{tab:top_tokens}
\small
\begin{tabular}{c|c|c|c}
\toprule
\textbf{Ticker} & \textbf{Transactions} & \textbf{Share (\%)} & \textbf{Protocol} \\
\midrule
\rowcolor{gray!10}
nodes & 1,084,718 & 22.8 & erc-20 \\
ethpi & 324,071 & 6.8 & erc-20 \\
\rowcolor{gray!10}
rETH & 277,183 & 5.8 & rerc-20 \\
eths & 187,279 & 3.9 & terc-20 \\
\rowcolor{gray!10}
ierc & 62,590 & 1.3 & ierc-20 \\
tom & 60,397 & 1.3 & terc-20 \\
\rowcolor{gray!10}
etfDay & 57,870 & 1.2 & erc-20 \\
powETH & 57,326 & 1.2 & ierc-pow \\
\rowcolor{gray!10}
terc & 50,612 & 1.1 & terc-20 \\
ieth & 49,858 & 1.0 & ierc-20 \\
\bottomrule
\end{tabular}
\end{table}

The $6.27$ million \Uone transactions contribute approximately $5.3$\,GB of permanent calldata to Ethereum's chain history. Because calldata cannot be pruned or deleted, this represents a perpetual storage cost on every full node.

\para{\Utwo Footprint} JSON operations are compact (median 75 bytes), so \Utwo contributes an estimated $\sim$$356$~MB ($4.75\text{M} \times 75\text{ bytes}$) of the total $5.3$\,GB footprint. While smaller than the media inscription footprint per transaction, \Utwo's volume ($75.8$\% of transactions) makes it the dominant contributor by transaction count, though a smaller contributor by byte volume (images and Facet operations carry larger payloads).

\subsubsection{Heavy-Hitter Contribution to Persistent Footprint}

The extreme sender concentration documented above translates directly into footprint inequality. The top $1{,}000$ senders ($0.5$\% of unique addresses) generate $34.1$\% of \Uone transactions. Assuming representative payload sizes, these heavy hitters are responsible for an estimated $\sim$$1.8$\,GB of the $5.3$\,GB permanent footprint. In the extreme tail, the single most active sender ($66{,}130$ transactions) contributes approximately $5$~MB of permanent calldata, replicated across every full node indefinitely. This inequality of \textit{footprint generation} (not just transaction counts) sharpens the externality argument: a small number of addresses impose disproportionate permanent storage costs on the entire network.

\subsubsection{Role Mix Over Time}

Self-send transactions (the canonical minting pattern) dominate throughout the active period, consistently comprising $>$$65$\% of monthly volume. Peer transfers remain a small fraction ($<$$5$\%), confirming that the \Utwo workload is overwhelmingly \textit{creation-oriented} rather than \textit{transfer-oriented}.

\begin{center}
\fbox{%
\begin{minipage}{0.9\linewidth}
\textbf{Takeaway:}
Extreme sender concentration (Gini~0.858) means a small fraction of addresses generates a disproportionate share of both transactions and the resulting 5.3\,GB permanent storage burden. Each inscriber pays a one-time gas fee that does not reflect the ongoing replication cost across every full node.
\end{minipage}}
\end{center}

\section{Discussion: System Implications}
\label{sec:discussion}

This section addresses \textbf{RQ4}: What design implications follow from the measured Ethscription workload?

\subsection{Persistent Footprint: What \Utwo Adds to Chain History}

The $5.3$\,GB of permanent calldata contributed by \Uone (with \Utwo comprising $75.8$\% of transactions) represents replicated history growth borne by every full node. Each inscriber pays a one-time gas fee that does not reflect the ongoing externality of permanent storage across the global node network. This is a direct instantiation of the tragedy of the commons~\cite{hardin1968tragedy, ostrom1990governing}: private benefit (permanent data storage) with distributed cost (node storage burden).

\subsection{Pricing vs.\ Permanence}

Calldata pricing (16~gas per non-zero byte) was calibrated for transient parameters, not permanent data embedding~\cite{wood2014ethereum}. Ethscription workload exposes a fundamental mismatch: the pricing model treats calldata as ephemeral execution input, yet the storage model makes it permanent. Design implications:

\begin{itemize}[leftmargin=*, nosep]
\item \textit{Explicit transient channels:} EIP-4844~\cite{buterin2022eip4844} blob storage ($18$-day expiry) addresses this for rollup DA workloads. Similar separation is needed for other data-intensive use cases.
\item \textit{Archival-burden pricing:} Pricing that reflects permanent replication cost would reduce the economic incentive for bulk calldata embedding.
\item \textit{Execution vs.\ embedding separation:} Protocol-level distinction between calldata for execution inputs and calldata for permanent data embedding.
\end{itemize}

\subsection{Indexer-Defined Semantics as a Software Layer}

The Ethscription ecosystem relies on off-chain indexers to derive token state from on-chain payloads (\Cref{sec:semantic_boundary}). This creates robustness risks:

\begin{itemize}[leftmargin=*, nosep]
\item \textit{Parser divergence:} Multiple indexer implementations may disagree on edge cases, leading to inconsistent balances across platforms.
\item \textit{Spec drift:} Protocol specification changes can retroactively alter historical state interpretation.
\item \textit{Replay/fork handling:} Chain reorganizations require indexers to re-derive state, with no guarantee of consistency.
\end{itemize}

\para{Empirical Evidence: Grammar-Rejected and Ambiguous Payloads}
Although \Utwo enforces a minimal grammar requiring the \texttt{p} and \texttt{op} fields, inspection of JSON-decoded payloads in \Uone reveals several classes of syntactically valid but semantically ambiguous submissions. These include: (i)~payloads missing required operational fields, (ii)~objects containing duplicated or conflicting keys (e.g., repeated \texttt{amt} or \texttt{tick}), and (iii)~numeric parameters encoded inconsistently (string vs.\ integer representations or leading-zero formats). Such cases constitute less than $0.1\%$ of \Uone, yet they are precisely the situations in which indexer implementations must apply local normalization or rejection policies. Representative patterns and normalization rules are summarized in Appendix~\ref{app:indexer_cases}.

For measurement, this means observable on-chain operations (\Utwo payloads) may not perfectly reflect semantic truth (what indexers actually accept). Our analysis measures the \textit{submitted} workload, not the \textit{validated} workload.

\subsection{Protocol Fragmentation Under Low Coordination Cost}

The $30$+ protocol identifiers (\Cref{tab:protocols}) with no convergence toward a single standard provide empirical evidence of fragmentation dynamics in permissionless environments. Unlike contract-based standards (\eg ERC-20~\cite{vogelsteller2015erc20}) where composability incentivizes convergence, inscription protocols have zero enforcement overhead: any participant can define a new \texttt{p} value. The result is persistent churn and non-convergence, where each new variant captures attention without bearing coordination costs.

This contrasts technically with smart contract standards, where on-chain enforcement (interface compliance, composability with DeFi protocols~\cite{werner2022sok,jiang2023decentralized}) creates network effects that reward standardization. The Ethscription ecosystem lacks these forcing functions.

\subsection{Practical Takeaways}

\para{For Protocol Designers} The Ethscription workload demonstrates that any permanent, underpriced data field will be exploited for data embedding. Future blockchain designs should separate transient execution data from permanent storage commitments with distinct pricing models.

\para{For the Measurement Community} The \Uzero/\Uone/\Utwo hierarchy illustrates the importance of explicit workload identification in blockchain measurement. Conflating Ethscriptions with general calldata (or conflating JSON operations with media inscriptions) produces misleading statistics. Scope-precise reporting is essential.

\begin{center}
\fbox{%
\begin{minipage}{0.9\linewidth}
\textbf{Takeaway:}
The Ethscription workload reveals a pricing–permanence mismatch: calldata, priced for transient execution, is used for permanent data storage at negligible cost. Protocol designers should distinguish transient execution data from permanent history growth, and measurement studies should use scope-aware datasets to avoid mixing workloads with different cost structures.
\end{minipage}}
\end{center}

\section{Limitations}
\label{sec:limitations}

\para{Prefix-Based Identification}
Our \Uone definition relies on the \texttt{data:} URI prefix as the primary identification signal. Transactions using non-standard encodings, indirect embedding methods, or payload constructions that omit this prefix are excluded. While this may undercount edge-case variants, the \texttt{data:} prefix constitutes the defining convention specified by Ethscriptions~\cite{lehman2023ethscriptions}, making our definition aligned with the intended canonical usage rather than heuristic inference.

\para{JSON Grammar Scope}
Our \Utwo grammar requires the presence of fields \texttt{p} and \texttt{op}, thereby excluding Ethscription variants adopting alternative or experimental schemas. This choice prioritizes semantic precision over maximal coverage. Sensitivity analysis under relaxed grammar constraints shows less than 0.1\% variation in \Utwo size, suggesting that excluded variants represent negligible ecosystem activity and do not materially affect aggregate observations.

\para{On-chain Only}
Our analysis considers only on-chain calldata payloads. Off-chain indexer interpretations, validated token balances, rejected operations, and marketplace-layer dynamics are intentionally excluded. As a result, our measurements characterize inscription behavior at the protocol interaction layer rather than economic outcomes or user-visible state maintained by external infrastructure.

\para{Concentration on \Uone}
Gini coefficients and Top-N concentration analyses are computed on \Uone rather than \Utwo alone, thereby including non-JSON transactions in concentration measurements. We state this explicitly to avoid scope ambiguity. Because \Utwo accounts for 75.8\% of \Uone and JSON minting dominates activity across both sets, the resulting concentration patterns are expected to remain structurally comparable, though minor deviations may exist at the margins.

\para{Partial February 2024}
Our dataset ends on February 26, 2024 (26 of 29 days). Although a declining trend is already observable, the final monthly aggregate is incomplete and may slightly underestimate total activity for the period. Interpretations involving month-level comparisons should therefore be viewed as provisional.

\para{Identity and Geography}
All analysis is conducted at the address level. We do not attempt to associate addresses with real-world identities, organizations, or geographic regions. Observed temporal signatures, including hourly activity patterns and lifecycle phases, are correlational signals only and should not be interpreted as demographic or geographic attribution.

\para{Correlation Not Causation}
Observed temporal and structural patterns reflect statistical association rather than causal inference. Our study identifies workload characteristics and ecosystem regularities but does not claim to establish causal mechanisms driving adoption or participation.

\textit{Comparison with Related Studies} \Cref{tab:comparison} positions our work against key related studies.

\begin{table*}[htb]
\centering
\caption{Comparison with related studies. \cmark~= full, \pmark~= partial, \xmark~= not addressed.}
\label{tab:comparison}
\small
\begin{tabular}{l |c c c c c c| c}
\toprule
\multicolumn{1}{c}{\textbf{Dimension}} &
Messias~\cite{writingonwall2024} &
Wang~\cite{wang2023brc20} &
Bertucci~\cite{bertucci2024ordinals} &
Li~\cite{li2024web3} &
Victor~\cite{victor2019measuring} &
Qi~\cite{qi2025understanding} & \textbf{Ours} \\
\midrule
Inscription-focused & \cmark & \cmark & \pmark & \cmark & \xmark & \xmark & \cmark \\
\rowcolor{gray!10}
Ethereum inscription ecosystem & \pmark & \xmark & \xmark & \pmark & \xmark & \xmark & \cmark \\
Scope-precise dataset hierarchy & \xmark & \xmark & \xmark & \xmark & \xmark & \xmark & \cmark \\
\rowcolor{gray!10}
Longitudinal ($>$6 months) & \xmark & \xmark & \pmark & \xmark & \cmark & \cmark & \cmark \\
JSON operation grammar & \xmark & \pmark & \xmark & \pmark & \xmark & \xmark & \cmark \\
\rowcolor{gray!10}
Protocol fragmentation analysis & \xmark & \pmark & \xmark & \pmark & \xmark & \cmark & \cmark \\
Lifecycle funnel metrics & \xmark & \pmark & \xmark & \pmark & \pmark & \xmark & \cmark \\
\rowcolor{gray!10}
Concentration (Gini) & \pmark & \xmark & \xmark & \xmark & \cmark & \xmark & \cmark \\
Persistent footprint & \pmark & \xmark & \cmark & \pmark & \xmark & \xmark & \cmark \\
\rowcolor{gray!10}
{Target chain} & Multi-EVM & Bitcoin & Bitcoin & Multi & Ethereum & Ethereum & Ethereum \\
\rowcolor{gray!10}
{Core analysis set} & All inscr. & BRC-20 & Ordinals & Survey & ERC-20 & NFT EIPs & \Utwo JSON \\
\bottomrule
\end{tabular}
\end{table*}

\section{Related Work}
\label{sec:related}

\para{Inscriptions and BRC-20s}
The inscription concept~\cite{li2024bitcoin} began with Bitcoin Ordinals~\cite{rodarmor2023ordinals}, enabling data attachment to individual satoshis via Taproot witness fields. Wang~\etal\cite{wang2023brc20} provided the first empirical analysis of BRC-20, examining the token mechanism, market performance, and user sentiment. Bertucci~\cite{bertucci2024ordinals} studied the determinants and impact of Bitcoin Ordinals on transaction fees. Li~\etal\cite{li2024web3} presented a study of inscription protocols as Web3 asset~\cite{wang2022exploring} innovation. Chen~\etal\cite{chen2024brc20pricing} examined BRC-20's impact on Bitcoin trading dynamics. Messias~\etal\cite{writingonwall2024} analyzed inscription activity across five EVM-compatible chains during a boom period, focusing on gas fee impacts. Beyond empirical studies, existing research has investigated broader dimensions of BRC-20 and inscription systems, particularly interoperability~\cite{wang2024bridging} and security implications~\cite{qi2024brc20,qi2025brc20}.

Our work differs in three ways: (i)~we focus exclusively on Ethereum Ethscriptions rather than Bitcoin inscription activity; (ii)~we provide a \textit{scope-precise} analysis targeting the JSON operational subset; and (iii)~we measure in a broad range covering workload dynamics, lifecycle mechanics, and protocol fragmentation, rather than primarily fee impacts.

\para{Ethereum calldata economics and data availability}
Calldata pricing in Ethereum was designed for smart contract parameters~\cite{wood2014ethereum, buterin2014whitepaper}. Alzoubi and Mishra~\cite{alzoubi2024blockchain} surveyed techniques to address blockchain bloat. EIP-4844~\cite{buterin2022eip4844} introduced blob transactions as an architectural response to calldata overuse. EIP-1559~\cite{buterin2019eip1559} reformed the fee market. Our work provides empirical evidence of a specific calldata overuse class (inscription workloads) and quantifies its persistent footprint.

\para{Blockchain measurements}
Large-scale empirical blockchain studies provide methodological precedent. Ron and Shamir~\cite{ron2013quantitative} pioneered quantitative Bitcoin transaction analysis. Meiklejohn~\etal\cite{meiklejohn2013fistful} characterized Bitcoin payment flows. Wang~\etal\cite{wang2023blockchain} studied censorship behaviours for on-chain transactions.Victor and L\"uders~\cite{victor2019measuring} measured ERC-20 token networks, finding concentration patterns.  Kim~\etal\cite{kim2018measuring} measured the Ethereum P2P network. Chen~\etal\cite{chen2020survey} surveyed Ethereum security. Wang~\etal\cite{wang2023snapshot} studied Snapshot DAO governance. Our concentration and temporal analysis methods follow this tradition.

\para{Contract standards vs.\ payload protocols}
The evolution of on-chain token standards~\cite{vogelsteller2015erc20} provides a technical contrast to inscription protocols. Qi~\etal\cite{qi2025understanding} analyzed NFT-related EIPs, finding poor cross-version interoperability even within formal standards. Contract-based standards enforce composability through on-chain interfaces; inscription protocols have no enforcement mechanism, producing the fragmentation we observe (\Cref{sec:proto_dynamics}).

\section{Conclusion}
\label{sec:conclusion}

We present a scope-precise empirical study of Ethereum Ethscriptions, focusing on the Ethscription operational subset (\Utwo). From 6.27M Ethscription candidates (\Uone), we extract 4.75M JSON-based Ethscription operations (75.8\%) and analyze their workload dynamics, protocol landscape, lifecycle mechanics, participation structure, and persistent footprint.

We deliver three design messages. First, calldata pricing that does not account for permanent replication cost creates an incentive for bulk data embedding. Protocol designers should separate transient data availability from permanent history growth. Second, the 30+ protocol variants with zero coordination overhead demonstrate that permissionless environments without enforcement mechanisms resist standardization. Third, indexer-defined semantics constitute a distinct software layer whose robustness properties (parser consistency, spec stability, fork handling) deserve first-class attention from both designers and the measurement community.

\section*{Responsible Measurement Note}
\label{app:responsible}

All data analyzed is publicly available on the Ethereum blockchain. We report only aggregate statistics (counts, distributions, ratios, Gini coefficients). Individual addresses are referenced only in aggregate form (``top 1,000 senders'') without linking to real-world identities, organizations, or geographic locations. Our analysis does not perform off-chain enrichment (ENS resolution, exchange attribution, IP geolocation). The pipeline is deterministic: given the same archive-node data and block range, all results are exactly reproducible.

\begin{table*}[htb]
\centering
\caption{Representative Ethscription edge cases that can trigger indexer divergence.}
\label{tab:indexer_edge_cases}
\small
\begin{tabular}{c| l}
\toprule
\textbf{Case type} &  \makecell{ \textbf{Example payload pattern} \&   (\text{why divergence may occur}) } \\
\midrule
Duplicate keys & \makecell{ \texttt{\{"p":"erc-20","op":"mint","tick":"x","amt":"10","amt":"1"\}} \\ JSON parsers vary: last-value-wins vs.\ reject; affects balance outcomes. }
\\

\rowcolor{gray!10}
Mixed numeric types & \makecell{
\texttt{\{"p":"erc-20","op":"mint","tick":"x","amt":001\}} \\
Leading zeros may be rejected by strict JSON; string-vs-int affects canonicalization.} \\

Whitespace / encoding &  \makecell{
\texttt{data:,\%7B"p":"erc-20",...\%7D} \\
URL-decoding rules and normalization differ; some indexers may treat as invalid.} \\

\rowcolor{gray!10}
Unknown fields &  \makecell{
\texttt{\{"p":"erc-20","op":"mint","tick":"x","amt":"10","foo":"bar"\}} \\
Spec interpretation: ignore unknown fields vs.\ reject to prevent spoofing.} \\

Case sensitivity &  \makecell{
\texttt{\{"p":"ERC-20","op":"Mint",...\}} \\
Whether protocol/op are case-normalized impacts matching and state updates. } \\
\bottomrule
\end{tabular}
\end{table*}

\bibliographystyle{unsrt}
\bibliography{references}
\appendices

\section{Extended Operation Details}

\subsubsection{Protocol Long-Tails}
\label{app:longtail}

Beyond the top 15 protocols reported in \Cref{tab:protocols}, the long tail includes 15+ additional protocol identifiers with fewer than 4,000 transactions each: \texttt{layer2-20} (3,705), \texttt{fair-20} (3,162), \texttt{eths} (2,906), \texttt{prc-20} (2,691), \texttt{bm-20} (2,509), \texttt{krc-20} (2,328), \texttt{eths-20} (2,227), \texttt{frc-20} (2,138), \texttt{erc20} (2,032), \texttt{base-20} (2,031), \texttt{eorc20} (1,831), \texttt{etw-20} (1,741), \texttt{derp-20} (1,592), \texttt{xrc-20} (1,569), \texttt{irc-20} (1,379).

Also, beyond the five lifecycle categories, additional operation types include: \texttt{pack} (10,136), \texttt{refund} (7,348), \texttt{stake} (6,855), \texttt{claim} (3,716), \texttt{unstake} (2,817), \texttt{bridge-recv} (2,649), \texttt{unfreeze\_sell} (2,153), \texttt{reg} (1,522), \texttt{unList} (1,400), \texttt{sellAck} (113), \texttt{split} (49), \texttt{send} (43).

\subsubsection{Indexer-Semantics Edge Cases \ref{tab:indexer_edge_cases}}
\label{app:indexer_cases}

Even with a minimal grammar, indexer-defined semantics can diverge on payload edge cases. We summarize representative cases that appear in \Uone JSON payloads and explain why they may yield inconsistent interpretations across indexers.

These edge cases are rare but critical: because Ethscription state is off-chain, any ambiguity in parsing or normalization can lead to inconsistent token balances across platforms. This motivates treating the indexer layer as a first-class software component with its own robustness and security properties.

\end{document}